\documentclass[graybox,vecphys]{svmult}

\usepackage{type1cm}        
\usepackage{makeidx}         
\usepackage{graphicx}        
\usepackage{multicol}        
\usepackage[bottom]{footmisc}

\usepackage{newtxtext}       %
\usepackage{newtxmath}       

\usepackage{cite}

\usepackage[pdftex, pdftitle={Article}, pdfauthor={Author}, colorlinks=true,linkcolor=blue,citecolor=blue,urlcolor=blue]{hyperref}
\usepackage[T1]{fontenc}
\usepackage{todonotes}
\usepackage{easyReview}

\makeindex             

\newcommand{\MLD}{\langle\mathrm{MLD}\rangle}
\newcommand{\ALD}{\langle\mathrm{ALD}\rangle}
\newcommand{\Rg}{R_{\mathrm{g}}}

\begin{document}

\title*{Viral RNA as a branched polymer}
\author{Domen Vaupoti\v{c}, Angelo Rosa, Rudolf Podgornik, Luca Tubiana, and An\v{z}e Bo\v{z}i\v{c}}
\authorrunning{Vaupoti\v{c} {\em et al.}}
\institute{Domen Vaupoti\v{c} \at Department of Theoretical Physics, Jo\v{z}ef Stefan Institute, Ljubljana, Slovenia
\and Angelo Rosa \at Scuola Internazionale Superiore di Studi Avanzati (SISSA), Trieste, Italy
\and Rudolf Podgornik \at School of Physical Sciences and Kavli Institute of Theoretical Science, University of Chinese Academy of Sciences, Beijing, China
\and Luca Tubiana \at Physics Department, University of Trento, Trento, Italy \at INFN-TIFPA, Trento Institute for Fundamental Physics and Applications, Trento, Italy
\and An\v{z}e Bo\v{z}i\v{c} \at Department of Theoretical Physics, Jo\v{z}ef Stefan Institute, Ljubljana, Slovenia \at \email{anze.bozic@ijs.si}}
%
%
\maketitle

\abstract{
Myriad viruses use positive-strand RNA molecules as their genomes. Far from being only a repository of genetic material, viral RNA performs numerous other functions mediated by its physical structure and chemical properties. In this chapter, we focus on its structure and discuss how long RNA molecules can be treated as branched polymers through planar graphs. We describe the major results that can be obtained by this approach, in particular the observation that viral RNA genomes have a characteristic compactness that sets them aside from similar random RNAs. We also discuss how different parameters used in the current RNA folding software influence the resulting structures and how they can be related to experimentally observable quantities. Finally, we show how the connection to branched polymers can be extended to take advantage of known results from polymer physics and can be further moulded to include additional interactions, such as excluded volume or electrostatics.
}


\section{Introduction}

RNA is an incredibly versatile biological macromolecule: not only does it act as a messenger between the DNA genome and the protein product, but it also assumes various roles in the form of transfer RNA, ribosomal RNA, microRNA, guide RNA, and long non-coding RNA, to name just a few~\cite{Eddy2001,Mattick2006}. Its function is carried out both on the level of its primary sequence of nucleotides and by the local and global structures that are formed when the constituent nucleotides form base pairs with each other~\cite{Gorodkin2014,Wang2021}. Many RNA structures are thus involved in translational control, RNA localization, gene regulation, RNA stability, and more~\cite{Mortimer2014}. RNA structure folding is hierarchical, with the formation of base pairs---described by {\em secondary} structure---dominating the contribution to the folding energy and leading into its embedding in three-dimensional space, described by {\em tertiary} structure~\cite{Brion1997,Mustoe2014}. In spite of recent improvements in the prediction of the tertiary structure of RNA molecules, it remains restricted to relatively short, individual sequences~\cite{Leontis2012,Miao2017,Li2021}. It is therefore of great advantage that RNA structure and its function can often be understood well by modelling it on the level of secondary structure, which can be further complemented by experimental methods such as SHAPE and its derivations~\cite{Low2010,Lorenz2016,Mitchell2019}.

In a large number of bacterial, plant, animal, and human {\em viruses}, positive-strand RNA (+ssRNA) takes on the role of their genomes~\cite{Holmes2009}. Far from simply coding for the protein products, both local structural elements as well as long-range structural interactions in the genomes of +ssRNA viruses are involved in many fundamental viral processes such as virus disassembly, translation, genome replication, and packaging, and are thus in general important for viral fitness~\cite{Liu2009,Newburn2015,Nicholson2015,Boerneke2019}. In particular, the genomes self-assemble together with capsid proteins to form a functional virion in an interplay of RNA sequence, length, and structure, further influenced by environmental variables such as pH and salt concentration~\cite{Schneemann2006,Rao2006,Garmann2016,ComasGarcia2019}. For instance, in certain viruses, local structural elements called packaging signals---typically one or several hairpin loops with a more or less defined structure and nucleotide pattern---are responsible for specific interactions with the capsid proteins, initiating assembly through several possible pathways~\cite{ComasGarcia2019,Twarock2018,Stockley2013}.

At the same time, non-specific electrostatic interactions between highly negatively charged RNA and positively charged domains of capsid proteins dominate the self-assembly of many +ssRNA viruses~\cite{Zandi2020,Perlmutter2015}. Here, a number of experiments have demonstrated that viral capsids can assemble not only with their native RNA genomes but also with non-cognate RNA genomes of other viruses, other RNA molecules, and even linear polyelectrolytes~\cite{Hu2008,ComasGarcia2012}. Success of the self-assembly and the resulting capsid(-like) structure, however, both depend on the {\em length and structure of the cargo} as well as on environmental variables~\cite{Beren2017,Marichal2021,Perlmutter2013}. Varying the salt concentration of the solution, for instance, changes the strength of RNA-protein interaction~\cite{Garmann2022}, and varying the strength of the interaction between RNA and an adsorbing substrate can change the latter's preference for adsorbing either single- or double-stranded RNA~\cite{Poblete2021}. 

The {\em branching structure of viral RNA}, in particular, plays an important role in RNA-capsid interaction and virus assembly. Experiments have demonstrated that RNA structure and topology influence both packaging efficiency and the resulting capsid size and shape~\cite{Beren2017,Marichal2021,Singaram2015}, while theoretical studies have shown that the degree of branching can greatly increase the amount of RNA that can be packaged into a capsid~\cite{Erdemci2014,Erdemci2016}. Moreover, branching patterns of different RNAs have been shown to influence their size~\cite{Gopal2014,Borodavka2016}, with genomes of +ssRNA viruses with icosahedral capsids being significantly more compact compared to those with helical capsids~\cite{Yoffe2008,Tubiana2015}---with the former capsid type providing more severe spatial restrictions than the latter. This characteristic compactness appears to be a global structural property, and while even $\sim 5\%$ of {\em synonymous} mutations were shown to destroy it~\cite{Tubiana2015}, the question remains of where in the genome sequence these topological and structural properties are encoded~\cite{Bozic2018}. Understanding the topological properties of the genomes of +ssRNA viruses is thus essential to understand their ability to self-assemble and consequently to design strategies to modify or interfere with their function~\cite{Farrell2022}.

In this chapter, we describe how the secondary structure of viral RNA can be mapped to a branched polymer, which properties can be extracted, what are some of the major results that can be obtained using this approach, and some pitfalls to be considered. To this purpose, we first introduce the main properties of branched polymers and demonstrate how RNA can be treated as one by being mapped onto a graph. We then describe some of the topological and structural properties that can be gleaned from this approach. Next, we illustrate this approach on random RNA sequences of different length and nucleotide composition, which provides a baseline for comparison of different biological RNAs. Focusing on the genomes of +ssRNA viruses, we explore the differences among them by comparing them to random RNAs as well as shuffled versions of themselves. We also show how model parameters used in the prediction of RNA secondary structure---specifically, multiloop energy and maximum base pair span---influence these predictions. Lastly, we briefly overview the field-theoretical description of RNA as a branched polymer, which makes use of the derived topological parameters and allows for a self-consistent inclusion of additional short- and long-range interactions in the analysis of interactions between the RNA genome and the capsid proteins.

\section{RNA as a branched polymer}
\label{sec:RNAbranch}

\subsection{Secondary structure of RNA as a graph}
\label{sec:graph}

\paragraph{{\em Secondary structure prediction}}

Description of RNA structure on the intermediate level of its secondary structure forms a conceptually important step and explains the dominant part of the free energy of structure formation~\cite{Fallmann2017}. Modelling RNA on this level allows for analysis of large numbers of very long RNA sequences---which would be prohibitively expensive to model on the level of their tertiary structure---while retaining the majority of the pertinent information about its local and global structure resulting from base-pairing. Numerous software packages exist for the prediction of RNA secondary structure, the most popular ones being ViennaRNA~\cite{Lorenz2011} and RNAstructure~\cite{Reuter2010}, based on energy models of base-pairing, and CONTRAfold~\cite{Do2006} and EternaFold~\cite{Wayment2020}, which learn model parameters using stochastic context-free grammar. All of these algorithms necessarily come with limitations~\cite{Wayment2020,Koodli2021,Liu2021}, but due to the complexity of structure prediction for long RNA molecules, they remain the tool of choice for studies of +ssRNA viral genomes, which can range anywhere from $\sim1000$--$30000$ nt in length (Sec.~\ref{sec:rnaVR}). While some of the uncertainty in the prediction of RNA secondary structure can be alleviated by taking into account experimental data~\cite{Lorenz2016}, such data is not widely available for most viral genomes.

Since the energy landscape of RNA structures is very shallow, predicting only the minimum free energy structure is typically insufficient, as the RNA can sample different conformations and several functional structures can co-exist in vivo~\cite{Spasic2018}. The benefit of using energy-based folding algorithms for the prediction of secondary structure is that they enable generation and sampling of {\em thermal ensembles} of representative structures at a given temperature~\cite{Mathews2006}. In the examples presented in this chapter, we use ViennaRNA v2.4~\cite{Lorenz2011} to predict thermal ensembles of $500$ structures at $T=37^\circ$~C for each RNA sequence and denote any quantity $\mathcal{O}$ averaged over this thermal ensemble of structures by $\langle\mathcal{O}\rangle$. As we show later on, this sample size produces sufficient statistics for each quantity we consider.

\begin{figure}[b]
\sidecaption[t]
\centering
\includegraphics[width=7.5cm]{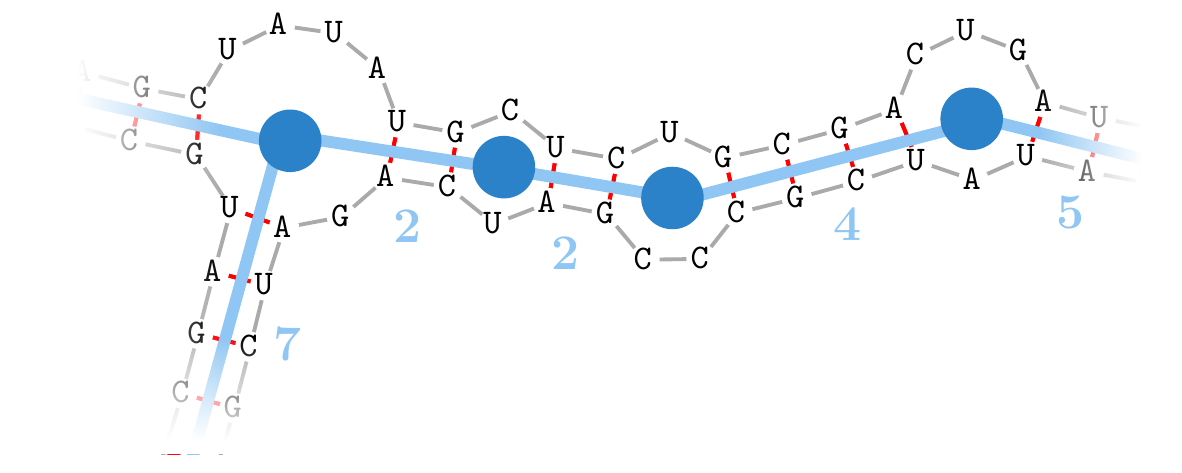}
\caption{Mapping (a part of) RNA secondary structure onto a planar tree. Double-stranded (base-paired) regions are mapped to graph edges, weighted with the stem length, while single-stranded regions are mapped to graph nodes.}
\label{fig:RAG}
\end{figure}

\paragraph{{\em RNA as a graph}}

The idea that the complexity of base pairs and sequence-structure patterns in a folded RNA sequence can be reduced by mapping its secondary structure onto a graph is not new (Ref.~\cite{Schlick2018} provides a detailed overview of the topic). In the absence of pseudoknots---a typical simplification which drastically reduces the computational complexity of structure prediction---the secondary structure can be described as a planar tree (Fig.~\ref{fig:RAG}). The simplest way to construct such a tree is by mapping double-stranded regions (base pairs) to {\em edges} with weights corresponding to the stem lengths, while single-stranded regions (unpaired nucleotides) are mapped to {\em nodes} connecting the edges. In the rest of the chapter, when we will refer to RNA trees, we will have in mind this procedure. This mapping is independent of the base-pairing model used to predict the structure---apart from the assumption of the absence of pseudoknots---and we discuss some differences that arise from using different model parameters in Sec.~\ref{sec:models}.

\begin{figure}[b]
\sidecaption[t]
\centering
\includegraphics[width=6.2cm]{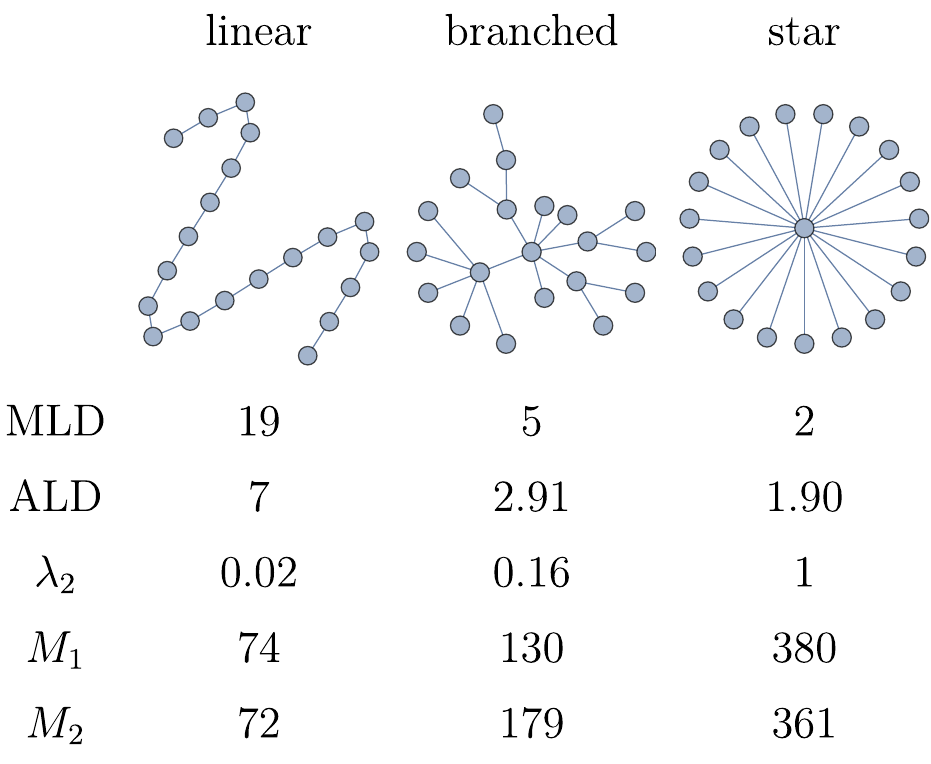}
\caption{Illustration of some topological quantities described in the main text: maximum (MLD) and average ladder distance (ALD), second Laplacian eigenvalue $\lambda_2$, and Zagreb indices $M_1$ and $M_2$. The values of these quantities are shown for three different types of polymers---linear, branched, and star polymer---all of them having the same total number of monomers $N$ and unit edge weight.}
\label{fig:comparison}
\end{figure}

Once RNA secondary structure is mapped onto a tree composed of $N+1$ nodes $v_{i}\in\mathbb{V}$, $i=0,\ldots,N$, connecting $N$ (undirected) edges $e_i=(v_j,v_k)\in\mathbb{E}$ with weights (stem lengths) $b_i$, it is possible to derive various parameters describing its topology and structure~\cite{GraphTheory,Todeschini2008,Rouvray2002}, including:
\begin{itemize}
    \item The distribution of ladder distances $p(\ell)$, where the ladder distance $\ell(v_i,v_j)$ is defined as the shortest path between a pair of nodes $v_i$ and $v_j$. The most important derived measures are the maximum ladder distance (MLD),
    \begin{equation}\label{eq:DefineMLD}
    \mathrm{MLD} = \max_{v_i,v_j\in\mathbb{V}}\ell(v_i,v_j),
    \end{equation}
    corresponding to the diameter of the graph, and the average ladder distance (ALD),
    \begin{equation}\label{eq:DefineALD}
        \mathrm{ALD} = \frac1{(N+1)N}\sum_{v_i\neq v_j\in\mathbb{V}}\ell(v_i,v_j),
    \end{equation}
    and its related quantity, the Wiener index $W=\sum_{v_i\neq v_j\in\mathbb{V}}\ell(v_i,v_j)$.
    \item The distribution of branch weights $p(N_\textrm{br})$, obtained by cutting the expanded tree at each edge and taking the smaller of the two total weights of the resulting trees.
    \item The distribution of node degrees $p(d_i)$, indicating the presence of multiloops (nodes of degree $d_i\geqslant 3$), with the total number of nodes with degree $k$ given by $D_k$. Some derived quantities are, e.g., Zagreb indices $M_1=\sum_{v_i\in\mathbb{V}}d_i^2$ and $M_2=\sum_{(v_i,v_j)\in\mathbb{E}}d_id_j$.
    \item The Laplacian spectrum, the eigenvalues $\lambda_i$ of the Laplacian matrix $\mathsf{L}=\mathsf{D}-\mathsf{A}$, where $\mathsf{D}$ is the matrix of node degrees and $\mathsf{A}$ is the node adjacency matrix. The second smallest eigenvalue $\lambda_2\leqslant 1$ describes the connectivity of the graph, with larger values indicating better connectivity or a more star-like structure.
\end{itemize}

\begin{figure}[hb]
\centering
\includegraphics[width=0.95\linewidth]{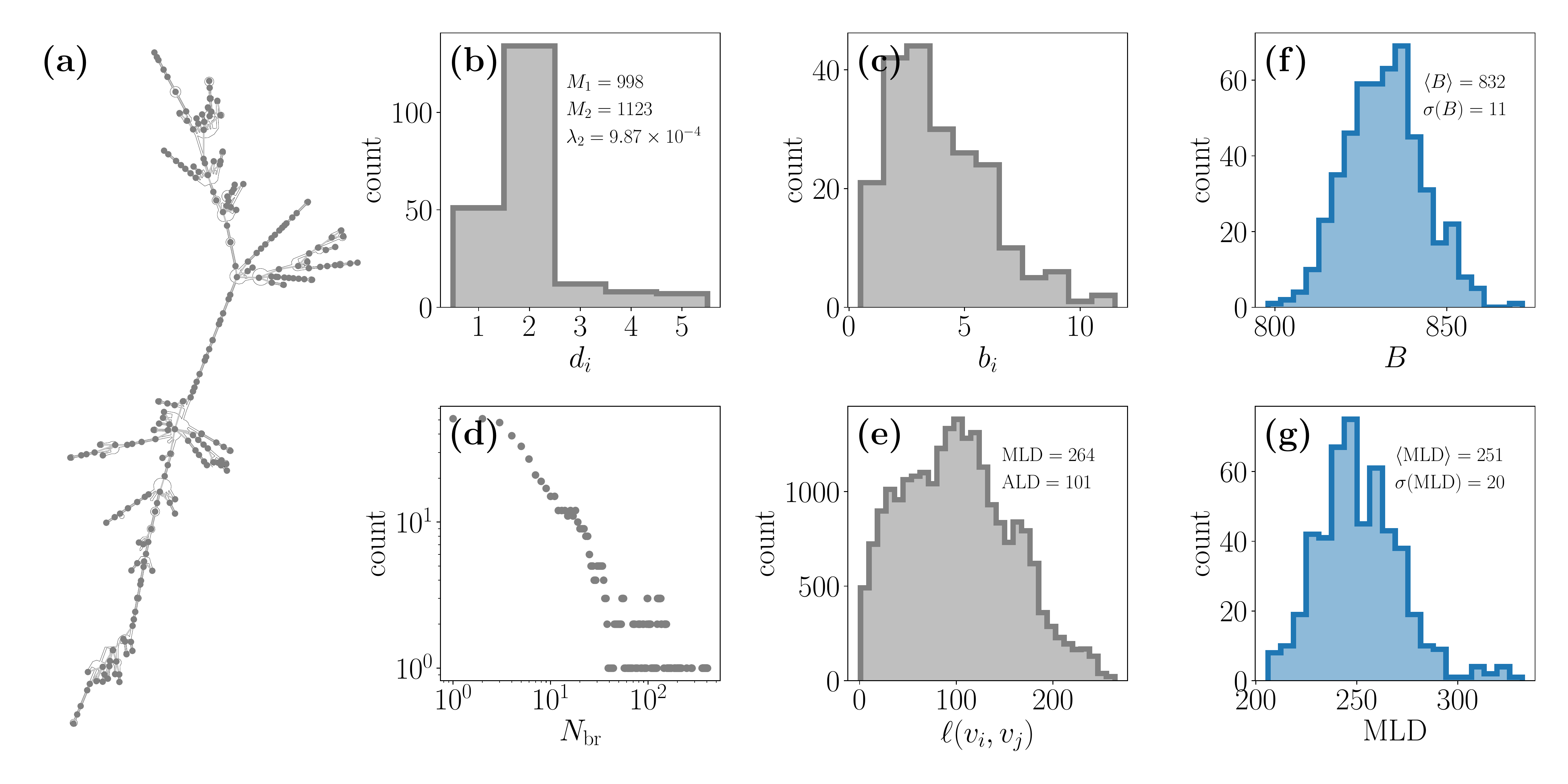}
\caption{Topological properties of RNA as a graph. {\bf (a)} Representative secondary structure of a random RNA ($N_\mathrm{nt}=2700$ nt), overlaid with its graph representation. {\bf (b)--(e)} Distributions of node degrees $d_i$, edge weights $b_i$, branch weights $N_\mathrm{br}$, and path lengths $\ell(v_i,v_j)$ for the RNA structure shown in panel (a). Also shown are the values of $M_1$ and $M_2$, $\lambda_2$, MLD, and ALD. {\bf (f)--(g)} Distributions of the total number of base pairs $B$ and of MLD for $500$ structures of the same RNA sequence drawn from a thermal ensemble. Also shown are the values of the thermal averages and their standard deviation.}
\label{fig:example}
\end{figure}

These quantities have the ability to distinguish, to various extents, between polymers with different types of tree topology, as illustrated in Fig.~\ref{fig:comparison}. Figure~\ref{fig:example} further illustrates some of these quantities on an example of a (uniformly) random RNA sequence, $N_\mathrm{nt}=2700$ nt in length. From the distribution of node degrees (panel (b)), one can for instance determine the Zagreb indices of the RNA tree, and from the distribution of ladder distances (panel (e)), one can determine both the MLD and the ALD. Panels (f) and (g) further show thermal ensemble distributions of MLD and the total number of base pairs $B$, demonstrating that their averages are well-defined quantities. As we will see in the following, combining different topological properties of RNA graphs with statistical mechanics of branched polymers can be used to gain insight into their physical properties. 

\subsection{Properties of branched polymers}
\label{sec:polymers}

Statistical mechanics of polymers is a very powerful theoretical tool with wide applications across biophysics, including, for instance, the scaling laws in the large-scale eukaryotic chromosome organization~\cite{Sazer2018} and liquid-liquid phase separation in cells~\cite{Perry2019}. An important conceptual peculiarity of systems of {\em branched} polymers (trees)---such as RNA---concerns the necessity to distinguish between {\em annealed} (or, randomly {\em branching})~\cite{Zandi2015} and {\em quenched} (or, randomly {\em branched}) polymers~\cite{Gutin1993}. Quenched trees are those whose topology of branches (tree connectivity) is fixed during the process of chemical synthesis and does not change afterwards. In contrast, the topology of annealed trees is not fixed at synthesis but instead can vary, typically in response to interactions (e.g., in the case of viral RNA, with capsid proteins) and/or changes in external conditions, and may fluctuate due to thermal motion. The class of annealed trees is particularly important as it is likely the most relevant for RNA molecules~\cite{Everaers2017}.

\paragraph{{\em Scaling exponents}}

A physical description of polymer conformations has to be formulated by adopting the probabilistic language of statistical mechanics. In particular, since the total number of accessible conformations of a polymer chain increases exponentially with the number $N$ of bonds (edges)~\cite{Wang2017}, polymers are best described in terms of averages of corresponding observables.

The most distinct feature of a polymer conformation is its linear size, which can be expressed in terms of the radius of gyration, defined as
\begin{equation}
\label{eq:Rg2-definition}
\Rg^2 \equiv \frac1N \sum_{i=1}^N (\vec r_i - \vec r_{\rm cm})^2,
\end{equation}
where
$\vec r_i$ is the spatial coordinate of the $i$-th monomer and $\vec r_{\rm cm} \equiv N^{-1} \sum_{i=1}^N \vec r_i$ is the centre-of-mass of the chain.
The characteristic mean polymer size (i.e., its {\em mean} gyration radius) is given by the square root of the statistical average of Eq.~\eqref{eq:Rg2-definition} over the ensemble of all accessible conformations,
\begin{equation}\label{eq:Rg-definition}
\langle \Rg(N) \rangle \equiv \sqrt{ \langle \Rg^2\rangle } \approx b N^\nu.
\end{equation}
The quantity $b$ is the mean bond length, while the {\em scaling exponent} $\nu$ is a fundamental parameter which---as we will shortly see---depends on several factors, particularly on monomer-monomer interactions~\cite{Giacometti2013}. While many fundamental works in polymer physics have dealt with determining the exponent $\nu$ for various polymer ensembles, exact values are known only for a very few cases~\cite{Everaers2017}. In most---and often the most relevant---cases, approximate (albeit accurate) results can be obtained by computationally extensive numerical methods or sophisticated mathematical tools~\cite{Everaers2017}. The values of $\nu$ for polymer ensembles most relevant in the context of RNA are shown in Table~\ref{tab:ScalingExps}; for other contexts, see the review by Everaers et al.~\cite{Everaers2017}.

\begin{table}[ht]
\caption{Best known values for scaling exponents of common polymer models in three dimensions.
Here, the ``ideal'' and ``self-avoiding'' refer to either a complete neglect or inclusion of excluded-volume effects, respectively (cf.\ Flory theory in Sec.~\ref{sec:polymers}). The values shown as fractions are exact, while others are approximate (obtained either from numerical simulations or by analytical methods). For linear polymers, we have trivially $\rho=\varepsilon=1$.}
\label{tab:ScalingExps}
\begin{tabular}{p{3.9cm}p{1.3cm}p{1.3cm}p{1.3cm}p{1cm}p{1cm}p{1.1cm}}
\hline\noalign{\smallskip}
Polymer model & $\nu$ & $\rho$ & $\varepsilon$ & $\nu_{\rm Flory}$ & $\rho_{\rm Flory}$ & Refs. \\
\noalign{\smallskip}\svhline\noalign{\smallskip}
Ideal linear & $1/2$ & $1$ & $1$ & $1/2$ & $1$ & \cite{RubinsteinColbyBook} \\
Self-avoiding linear & $0.5877$ & $1$ & $1$ & $3/5$ & $1$ & \cite{LiMadrasSokal1995} \\
Ideal branching & $1/4$ & $1/2$ & $1/2$ & $1/4$ & $1/2$ & \cite{RubinsteinColbyBook} \\
Self-avoiding branching & $1/2$ & $0.654$ & $0.651$ & $7/13$ & $9/13$ & \cite{ParisiSourlas1981,vanRensburg1992} \\
\noalign{\smallskip}\hline\noalign{\smallskip}
\end{tabular}
\end{table}

While the exponent $\nu$ is sufficient to understand the physical properties of {\em linear} polymers (see Fig.~\ref{fig:comparison}), to completely understand an ensemble of {\em branching} polymers, such as viral RNAs, it is also necessary to characterize the topology of branching (or, equivalently, the {\em tree connectivity})~\cite{Everaers2017}. This is a particularly central problem for RNA secondary structure, since its mean gyration radius (Eq.~\eqref{eq:Rg-definition}) and hence the exponent $\nu$ are not easily accessible.

The problem of characterizing the connectivity of various ensembles of branching polymers has been theoretically addressed numerous times~\cite{vanRensburg1992,RosaEveraersJPA2016,RosaEveraersJCP2016}. In the context of RNA, it is useful to introduce as a proper measure of chain connectivity the ensemble average of either the MLD or the ALD (Eqs.~\eqref{eq:DefineMLD} and~\eqref{eq:DefineALD}) as a function of the number of monomers $N$,
\begin{equation}
\label{eq:L-definition}
\langle {\rm MLD}(N) \rangle \sim \langle {\rm ALD}(N) \rangle \sim b N^\rho ,
\end{equation}
both of which account for the average length of linear paths on the tree.
While originally defined for characterizing the connectivity and introduced independently from $\nu$, the exponent $\rho$ in Eq.~\eqref{eq:L-definition} is related to it, and consequently also provides a fundamental insight into RNA folding {\em in physical space}.

Last but not least, we can also consider the average branch weight~\cite{vanRensburg1992}:
\begin{equation}\label{eq:Nbr-definition}
\langle N_{\rm br}(N) \rangle \sim N^\varepsilon ,
\end{equation}
which is defined as the average weight of the smallest of the two sub-trees obtained by systematically removing---one at time---the edges connecting two neighbouring nodes of the original tree~\cite{RosaEveraersJPA2016}. Note that while the two scaling exponents $\rho$ and $\epsilon$ describe very different quantities, they are not independent from each other. In fact, the relation
\begin{equation}\label{eq:EpsilonEqualRho}
\varepsilon = \rho    
\end{equation}
is expected to hold for randomly branching polymers in general~\cite{vanRensburg1992}. Equation~\eqref{eq:EpsilonEqualRho} is particularly appealing because it can be used to support {\em a posteriori} the initial hypothesis that RNA behaves as a randomly branching polymer: In fact, it is ``sufficient'' to measure $\ALD$ and $\langle N_{\rm br}\rangle$ as a function of $N$ and compare the estimates for the corresponding scaling exponents. Table~\ref{tab:ScalingExps} again summarizes the known values of $\rho$ and $\varepsilon$ for selected polymer ensembles.

Even the simplest theory of branching polymers thus has to deal with the three distinct observables introduced in Eqs.~\eqref{eq:Rg-definition}--\eqref{eq:Nbr-definition}. Since we are primarily interested in the secondary structure of random and viral RNAs, we will focus on topological observables such as $\MLD$ and $\langle N_{\rm br} \rangle$. Nonetheless, we will show that this has important consequences for how RNA molecules fold in space, i.e., on the average molecular size as given by $\langle \Rg \rangle$.

\paragraph{{\em Flory theory}}

Exact values for the scaling exponents $\nu$ and $\rho$ are known only in few special polymer cases. In this respect, Flory theories of polymers~\cite{FloryChemBook,Giacometti2013,Everaers2017} provide a simple framework for first---and yet remarkably accurate---approximations of both $\nu$ and $\rho$. Flory theory is formulated in terms of a balance between {\em (i)} an entropic (elastic) term, given by a sum of two contributions coming from the classical entropy of swelling ($F_{\rm sw}$) and the entropy of reconfiguration of the tree architecture ($F_{\rm tree}$) due to swelling and interaction, and {\em (ii)} an interaction term ($F_{\rm inter}$) arising from monomer-monomer collisions. Taken together, the Flory free energy (in units of $\beta^{-1}=k_BT$) reads~\cite{FloryChemBook,Giacometti2013,Everaers2017}:
\begin{eqnarray}\label{eq:Flory}
    F
    & = & F_{\rm sw}(N, \langle \Rg\rangle, \langle {\rm ALD}\rangle) + F_{\rm tree}(N, \langle {\rm ALD}\rangle) + F_{\rm inter}(N, \langle \Rg\rangle) \nonumber\\
    & = & \frac{\langle \Rg\rangle^2}{\langle {\rm ALD}\rangle b} + \frac{\langle {\rm ALD}\rangle^2}{Nb^2} + \upsilon_2 \frac{N^2}{\langle \Rg\rangle^3} ,
\end{eqnarray}
where $\upsilon_2\sim b^3$ is on the order of the {\em second virial coefficient}~\cite{Everaers2017}, accounting for the excluded-volume interaction between any two monomers. Although physically appealing, this representation of the free energy is itself an approximation, since the terms in the free energy are not independent from one another. Nonetheless, Flory theories turn out to be quite accurate~\cite{FloryChemBook,Giacometti2013,Everaers2017}.

A key feature of the free energy in Eq.~\eqref{eq:Flory} is that the interaction term does not depend on $\ALD$, which likely remains valid even in other ensembles with different forms of interaction energy $F_{\rm inter}$~\cite{Everaers2017}.
Consequently, we can balance the first two terms without worrying about the third, and connect $\langle \Rg\rangle$, $\langle{\rm ALD}\rangle$ and $N$: 
\begin{equation}\label{eq:LvsR}
    \langle {\rm ALD}\rangle \sim b^{1/3} N^{1/3} \langle \Rg\rangle^{2/3} ,
\end{equation}
or, conversely (see Eqs.~\eqref{eq:Rg-definition} and~\eqref{eq:L-definition}),
\begin{equation}\label{eq:RhovsNu}
\rho = \frac{1+2\nu}3 \Longleftrightarrow \nu = \frac{3\rho-1}2 .
\end{equation}
By reinserting Eq.~\ref{eq:RhovsNu} into Eq.~\eqref{eq:Flory} and balancing the remaining terms, we finally get the estimates for $\nu$ and $\rho$ shown in Table~\ref{tab:ScalingExps}, which, when compared to the exact ones, are remarkably accurate. In general, the relation between $\nu$ and $\rho$ in Eq.~\eqref{eq:RhovsNu} has been compared in various ensembles of randomly branching polymers and has been found to be very accurate in all cases~\cite{Everaers2017}. Its practical implications are quite remarkable, as it allows us to connect branching ($\rho$) to $3D$ conformations ($\nu$) by determining either of the two exponents in terms of the other. In the context of RNA, this relation is particularly useful, since we can determine $\rho$ from the topological properties of its structure and extract $\nu$ afterwards.

\section{Branching properties of viral RNAs}
\label{sec:rnaVR}

\subsection{Random RNAs}

Unlike viral RNA genomes which have a well-defined sequence length, random RNAs can be used to generate sequences of (in principle) arbitrary length and nucleotide composition. This enables one to explore how their topological properties change with both length and composition and in this way obtain different scaling relationships (Sec.~\ref{sec:polymers}). It is important to note here that RNA sequence length $N_\mathrm{nt}$ is, on average, directly proportional to the tree size $N$ of its structure, and the two quantities can be used interchangeably.

\begin{figure}[b]
\centering
\includegraphics[width=0.95\linewidth]{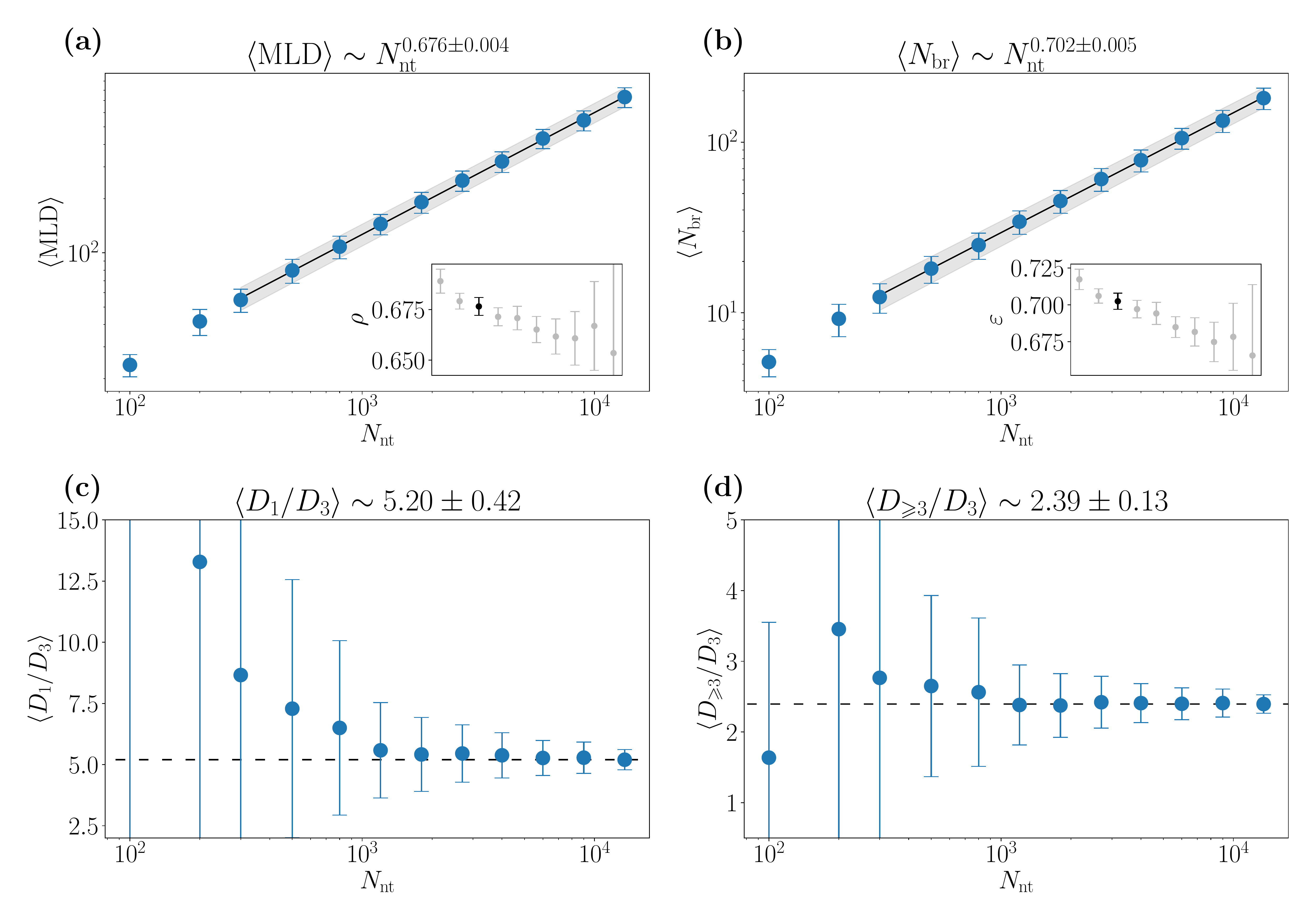}
\caption{Scaling of topological properties of (uniformly) random RNA sequences with their length: {\bf (a)} maximum ladder distance $\MLD$, {\bf (b)} branch weight $\langle N_\mathrm{br}\rangle$, {\bf (c)} ratio of the number of degree $1$ and degree $3$ nodes $D_1/D_3$, and {\bf (d)} ratio of the number of degree $3$ and all multiloop ($\geqslant3$) nodes, $D_{\geqslant 3}/D_3$. Each point in the plots represents an average over $200$ random sequences and the error bars show the standard deviation. Insets in panels (a) and (b) show how the scaling exponents change with the starting point of the fit, with the black point corresponding to a fit over the shaded region in the panel.}
\label{fig:randlen}
\end{figure}

Figure~\ref{fig:randlen} shows some examples of scaling laws for uniformly random RNA sequences, $f(\mathrm{A})=f(\mathrm{C})=f(\mathrm{U})=f(\mathrm{G})=0.25$, where $f(n)$ is the frequency of a nucleotide in the sequence. While some properties, such as $\MLD$ and $\langle N_\mathrm{br}\rangle$, follow a scaling law with a well-defined exponent, others, such as the ratio of the number of degree $1$ nodes (leaves of the tree---corresponding to hairpin configurations of RNA) and degree $3$ nodes, $D_1/D_3$, tend towards a constant value. The scaling exponents are of course {\em asymptotic} properties valid for large RNA structures, as seen in the insets in panels (a) and (b) of Fig.~\ref{fig:randlen}, which show how the fitted values of exponents change as shorter sequences are progressively removed from the fit.

\begin{figure}[ht]
\centering
\includegraphics[width=0.95\linewidth]{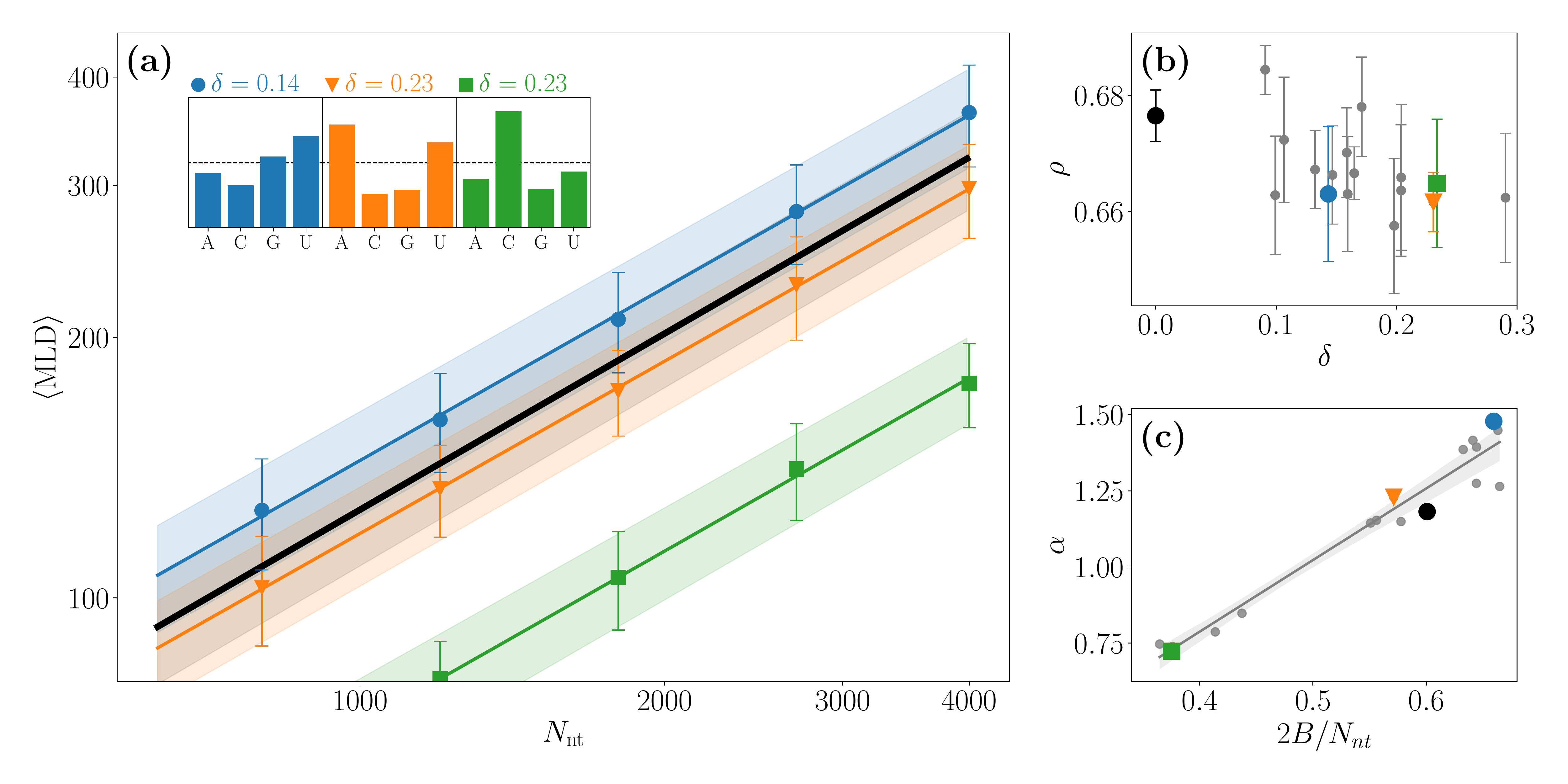}
\caption{
{\bf (a)} Scaling of $\MLD$ with sequence length, $\MLD=\alpha N_\mathrm{nt}^\rho$, for random RNA sequences with different nucleotide compositions. Each point in the plots represents an average over $200$ random sequences with $500$ thermal ensemble folds for each.
{\bf (b)} Scaling exponent $\rho$ for random RNAs as a function of the Euclidean distance from the uniform composition $\delta$.
{\bf (c)} Scaling prefactor $\alpha$ for RNAs with different compositions as a function of the base pair percentage, $2B/N_\mathrm{nt}$.}
\label{fig:randcomp}
\end{figure}

Nucleotide composition can vary significantly between different viral species (and biological RNAs in general)~\cite{Simon2021,Schultes1997}, affecting their properties. Still, changing the composition of random RNA sequences mainly influences the prefactor of the $\MLD$ scaling law (Fig.~\ref{fig:randcomp}a) and only minimally its exponent (Fig.~\ref{fig:randcomp}b), even when their composition deviates significantly from a uniformly random one, as evaluated by the Euclidean distance $\delta^2=\sum_{n\in\{\mathrm{A,C,G,U}\}}[f(n)-0.25]^2$. The change in the prefactor appears to be related to a decrease in the base pair percentage $2B/N_\mathrm{nt}$ (Fig.~\ref{fig:randcomp}c), which is perhaps unsurprising, as this leads to a smaller size of the RNA graph $N$ at the same sequence length $N_\mathrm{nt}$.

\subsection{Viral RNAs}

\begin{figure}[t]
\centering
\includegraphics[width=0.95\linewidth]{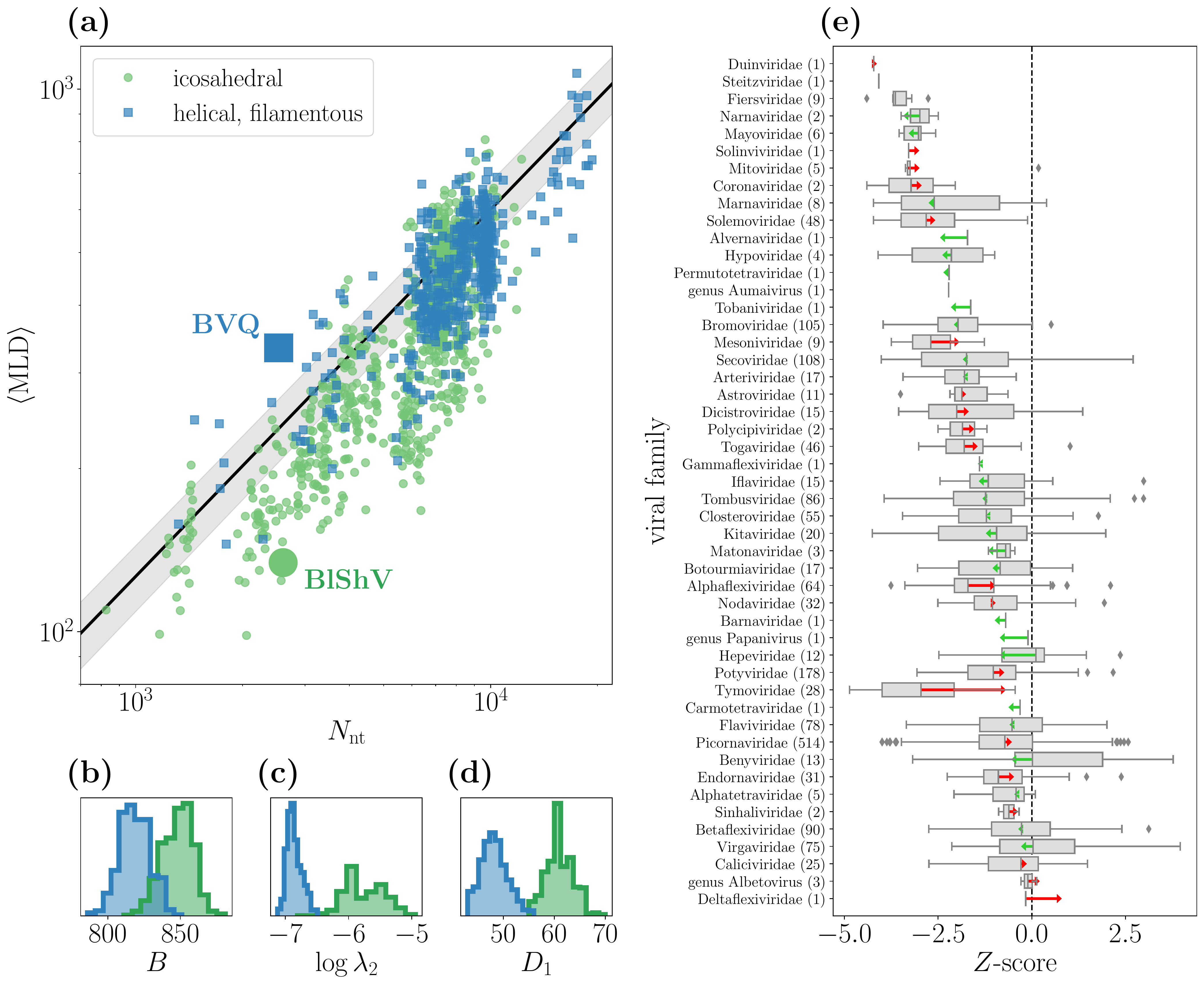}
\caption{{\bf (a)} $\MLD$ of $\sim 1500$ genomes of +ssRNA viruses of different lengths and capsid types. Black line shows the scaling for random RNA sequences with uniform composition, $\MLD\sim N_\mathrm{nt}^{0.676}$, and the shaded area shows the region where $|Z|\leqslant1$. {\bf (b)--(d)} Number of base-pairs $B$, logarithm of the second Laplacian eigenvalue $\log\lambda_2$, and number of degree $1$ nodes $D_1$ of genomes of two viruses indicated in panel (a), BQV (large square) and BlShV (large circle). {\bf (e)} Distribution of $\MLD$ $Z$-scores in different viral families, calculated with respect to random RNA sequences with uniform composition. Arrows indicate the shift in the median $Z$-score when the $\MLD$ is calculated with respect to random RNA with closest viral-like composition. The number of genomes included in each viral family is noted in parentheses next to its name.}
\label{fig:mld}
\end{figure}

Random RNA sequences show what we can expect of biological and viral RNAs in general~\cite{Higgs1993,Clote2005}. An important example of this is the compactness of the viral RNA folds as captured by its proxy measure, the MLD (cf.\ Sec.~\ref{sec:polymers}). As already mentioned, the MLD of the genomes of +ssRNA viruses with icosahedral capsids, which need to pack the genome into a comparatively small volume, was found to be significantly smaller compared to random RNA sequences of viral-like composition~\cite{Yoffe2008,Tubiana2015}. On the other hand, the MLD of genomes of viruses with helical capsids, which can in principle extend indefinitely, was indistinguishable from that of random RNAs. Figure~\ref{fig:mld}a demonstrates these differences on an extended set of $\sim1500$ genomes of +ssRNA viruses from different families, obtained from the Virus Metadata Resource of ICTV~\cite{Lefkowitz2018}, highlighted by capsid type.

Such an analysis opens up the possibility of comparing other topological properties of compact and non-compact viruses, as illustrated in panels (b)--(d) of Fig.~\ref{fig:mld} for genomes of beet virus Q (BVQ) and blueberry shock virus (BlShV), which are of comparable length but significantly less and more compact than random RNA, respectively (Fig.~\ref{fig:mld}a). The second Laplacian eigenvalue, for instance, identifies the compact genome of BlShV as more star-like (cf.\ Fig.~\ref{fig:comparison}), but even more interesting is that the genome of BlShV forms {\em more} base pairs compared to the genome of BVQ and is at the same time located {\em lower} than the scaling law for uniformly random RNAs. This is in direct contrast to observations in random RNAs with different nucleotide composition, where the scaling prefactor was reduced for those RNAs which form fewer base pairs (Fig.~\ref{fig:randcomp}), and indicates that the compactness of viral RNAs goes beyond simple differences in nucleotide composition.

Difference in a quantity $\mathcal{O}$ between viral and random RNAs can also be evaluated through the $Z$-score,
\begin{equation}
Z=\frac{\langle \mathcal{O}\rangle_\textrm{viral}-\langle \mathcal{O}\rangle_\textrm{random}}{\sigma(\mathcal{O})_\textrm{random}}.
\label{eq:zscore}
\end{equation}
As Fig.~\ref{fig:mld}e shows, this allows to study the properties of different sets of genomes, in this case grouped by viral family. It is immediately obvious that genomes in certain families are overall more compact than what would be expected of similar random RNAs, while the compactness of genomes in other viral families is indistinguishable from random RNAs. At the same time, there is also quite some degree of variation within viral families. Importantly, the difference in compactness typically persists no matter whether the genome $\MLD$ is compared to random RNA with a nucleotide composition similar to the one of the genome or to a uniformly random RNA, as indicated by the arrows in Fig.~\ref{fig:mld}b. (Notable exception are Tymoviridae, which have a significantly different composition~\cite{Yoffe2008,Tubiana2015}.) This implies that the difference in nucleotide composition of various +ssRNA genomes does not suffice to explain the resulting differences in their compactness as measured by the $\MLD$.

\paragraph{{\em $k$-let shuffle of viral genomes}}

Both mononucleotide and dinucleotide frequencies of viral RNA genomes exhibit biases among different viral species~\cite{Simon2021,Gaunt2022}, even if they share the same host~\cite{Giallonardo2017}. These biases are reflected in other properties---for instance, dinucleotide frequencies at codon position $2$-$3$ were shown to explain the majority of codon usage bias~\cite{Belalov2013}. Studies have made it clear that nucleotide composition alone does not suffice to explain the observed differences in the $\MLD$ of viral genomes~\cite{Yoffe2008,Tubiana2015} (see also Fig.~\ref{fig:mld}), which is further supported by computational observations that {\em synonymous} mutations preserving both mononucleotide and dinucleotide frequencies easily erase their characteristic compactness~\cite{Tubiana2015,Bozic2018}.

\begin{figure}[ht]
\centering
\includegraphics[width=0.95\linewidth]{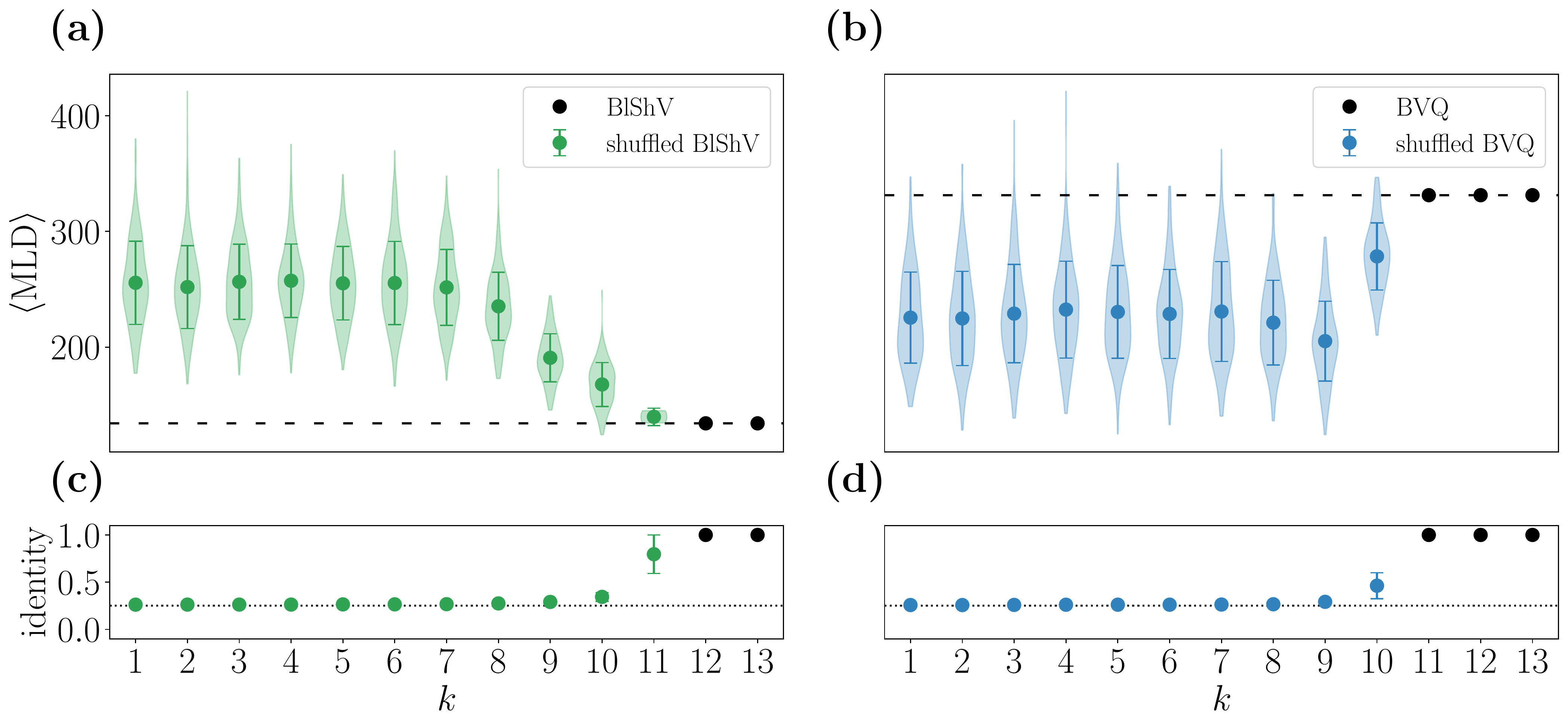}
\caption{$\MLD$ of shuffled sequences of {\bf (a)} BlShV and {\bf (b)} BVQ genomes with preserved $k$-let frequencies and {\bf (c)} and {\bf (d)} their corresponding sequence identities. The genomes of these two viruses are more and less compact than comparable random RNA, respectively (Fig.~\ref{fig:mld}a). The distributions represent sequence ensembles, with $200$ shuffled sequences used where possible (i.e., except for $k\geqslant11$ and the wild-type genome).}
\label{fig:kshuffle}
\end{figure}

One can thus take a step further and compare instead the topological properties of viral RNAs with their shuffled versions which conserve higher-order nucleotide frequencies of the original genomes. This can be achieved by using a $k$-let preserving shuffling algorithm, such as implemented by uShuffle~\cite{Jiang2008}, to shuffle the original genome sequences while {\em exactly} preserving $k$-let nucleotide frequencies. This means that for $k=1$ we preserve mononucleotide frequencies (nucleotide composition), for $k=2$ we preserve dinucleotide frequencies, and so on. Of course, as $k$ increases, the number of possible shuffled sequences decreases. In the examples of BVQ and BlShV, there are still $\sim10^3$--$10^4$ possible shuffled sequences available for $k=10$, while the count drastically drops for $k=11$ where only $\lesssim 10$ different shuffled sequences exist (Fig.~\ref{fig:kshuffle}). For the non-compact genome of BVQ, $k$-let shuffling does not produce drastic changes and the $\MLD$ of the shuffled sequences is comparable to that of random RNA for all possible values of $k$ (Fig.~\ref{fig:kshuffle}b). On the other hand, for the compact genome of BlShV, there is a drastic change in the range of $k=8$--$10$. For lower values of $k$, the shuffle completely destroys the compactness of the genome, as was previously seen for $k=1$ and $k=2$~\cite{Yoffe2008,Tubiana2015,Bozic2018}. For $k=9$, however, the $\MLD$ remains close to the compactness of the original genome, and for $k=10$ it is indistinguishable from it. There thus appears to be a particular length scale ($k\approx10$) at which shuffling the compact viral genomes while preserving their $k$-let nucleotide frequency also preserves their compactness. This observation could hold important clues to the question of where at the sequence level the genome compactness is encoded, but of course needs to be explored more carefully.

\section{Influence of model parameters}
\label{sec:models}

RNA secondary structure treated as a branched polymer provides a lot of information about the topological and structural nature of +ssRNA viral genomes and random RNAs alike. The inability to exactly predict the base pair patterns in a folded RNA sequence, however, can lead to differences in the predicted topology of the secondary structure, with consequences also for the subsequent tertiary structure prediction~\cite{Zhao2018}. While prediction of thermal structure ensembles necessitates the use of energy-based models, there is nonetheless a wide variety of model parameters that can influence the resulting predicted structures. We will briefly comment on the effect that two of the most important ones---namely, energy of multiloop formation and the maximum allowed base pair span---have on the topological measures of secondary structure of viral RNA genomes.

\subsection{Multiloop energy models}
\label{ssec:energy}

Energy-based folding algorithms predict both the minimum free energy fold of an RNA sequence as well as its pairing probability matrix, from which an ensemble of thermal folds can be obtained. These algorithms typically use a nearest neighbour energy model that breaks down the energy of an RNA structure into a sum of energies of its constituent loops. Commonly used sets of energy parameters are based on measurements provided by Turner~\cite{Turner2010}, with two particular sets of parameters---Turner1999 and Turner2004---used as a basis by different versions of the most popular energy-based folding software such as ViennaRNA and RNAstructure. Several efforts have also been made to improve on these parameter sets by using various computational optimization techniques~\cite{Andronescu2010,Langdon2018}.

Among the numerous energy parameters involved in structure prediction, {\em multiloop energies} are the least accurately known~\cite{Poznanovic2021}, even though occurrences of multiloops of degree $10$ or higher are not uncommon in various RNAs~\cite{Wiedemann2022}. Since allowing for an arbitrary size of a multiloop increases the computational complexity of structure prediction, earliest energy-based structure prediction models simply neglected multiloop contributions to the energy~\cite{Zuker1981}. Most of the current structure prediction software assumes that the energy of a multiloop depends only on the amount of enclosed base pairs (number of branches) and the number of unpaired nucleotides in it, and uses a linear model of the form
\begin{equation}
\label{eq:multiloop}
E_\mathrm{multiloop}=E_0+E_\mathrm{br}\times[\mathrm{branches}]+E_\mathrm{un}\times[\textrm{unpaired nucleotides}],
\end{equation}
where $E_0$ is the energy contribution for multiloop initiation, and $E_\mathrm{br}$ and $E_\mathrm{un}$ are the energy contributions for each enclosed base pair and unpaired nucleotide, respectively. While this form has been chosen mostly for its computational simplicity, improved models of multiloop energy that have been proposed seemingly do not lead to improved multiloop predictions compared to the linear one~\cite{Ward2017,Ward2019}.

\begin{table}[b]
\sidecaption
\begin{tabular}{p{3.7cm}p{0.8cm}p{0.8cm}p{0.8cm}p{0.6cm}}
\hline\noalign{\smallskip}
Energy model & $\phantom{+}E_0$ & $\phantom{+}E_\mathrm{un}$ & $\phantom{+}E_\mathrm{br}$ & Ref.  \\
\noalign{\smallskip}\svhline\noalign{\smallskip}
Turner1999 & $10.1$ & $-0.3$ & $-0.3$ & \cite{Turner2010} \\
Turner2004 & $\phantom{0}9.25$ & \phantom{+}NA & $\phantom{+}0.63$ & \cite{Turner2010} \\
ViennaRNA ($<$ v2.0) & $\phantom{0}3.4$ & $\phantom{+}0.0$ & $\phantom{+}0.4$ & \cite{Lorenz2011} \\
ViennaRNA (v2.0$+$) & $\phantom{0}9.3$ & $\phantom{+}0.0$ & $-0.9$ & \cite{Lorenz2011} \\
RNAstructure & $\phantom{0}9.3$ & $\phantom{+}0.0$ & $-0.6$ & \cite{Reuter2010} \\
Andronescu2007 & $\phantom{0}4.4$ & $\phantom{+}0.04$ & $\phantom{+}0.03$ & \cite{Andronescu2010} \\
Langdon2018 & $\phantom{0}9.3$ & $\phantom{+}0.0$ & $-0.8$ & \cite{Langdon2018} \\
\noalign{\smallskip}\hline\noalign{\smallskip}
\end{tabular}
\caption{Comparison of different energy parameters for the linear multiloop energy model [Eq.~\eqref{eq:multiloop}] used in prediction of RNA secondary structure.}
\label{tab:multiloop}
\end{table}

Table~\ref{tab:multiloop} gives an overview of some of the most commonly used energy parameters for the linear multiloop model (Eq.~\eqref{eq:multiloop}). A notable difference between the models lies not only in the magnitude but in {\em the sign} of the parameter $E_\mathrm{br}$ which controls the number of branches stemming from the multiloop. Earlier versions of ViennaRNA (until v2.0), for instance, used a positive value of this parameter, penalizing high-degree nodes, while the latest versions of the software use a negative value, promoting high-degree nodes.

\begin{figure}[t]
\centering
\includegraphics[width=0.95\linewidth]{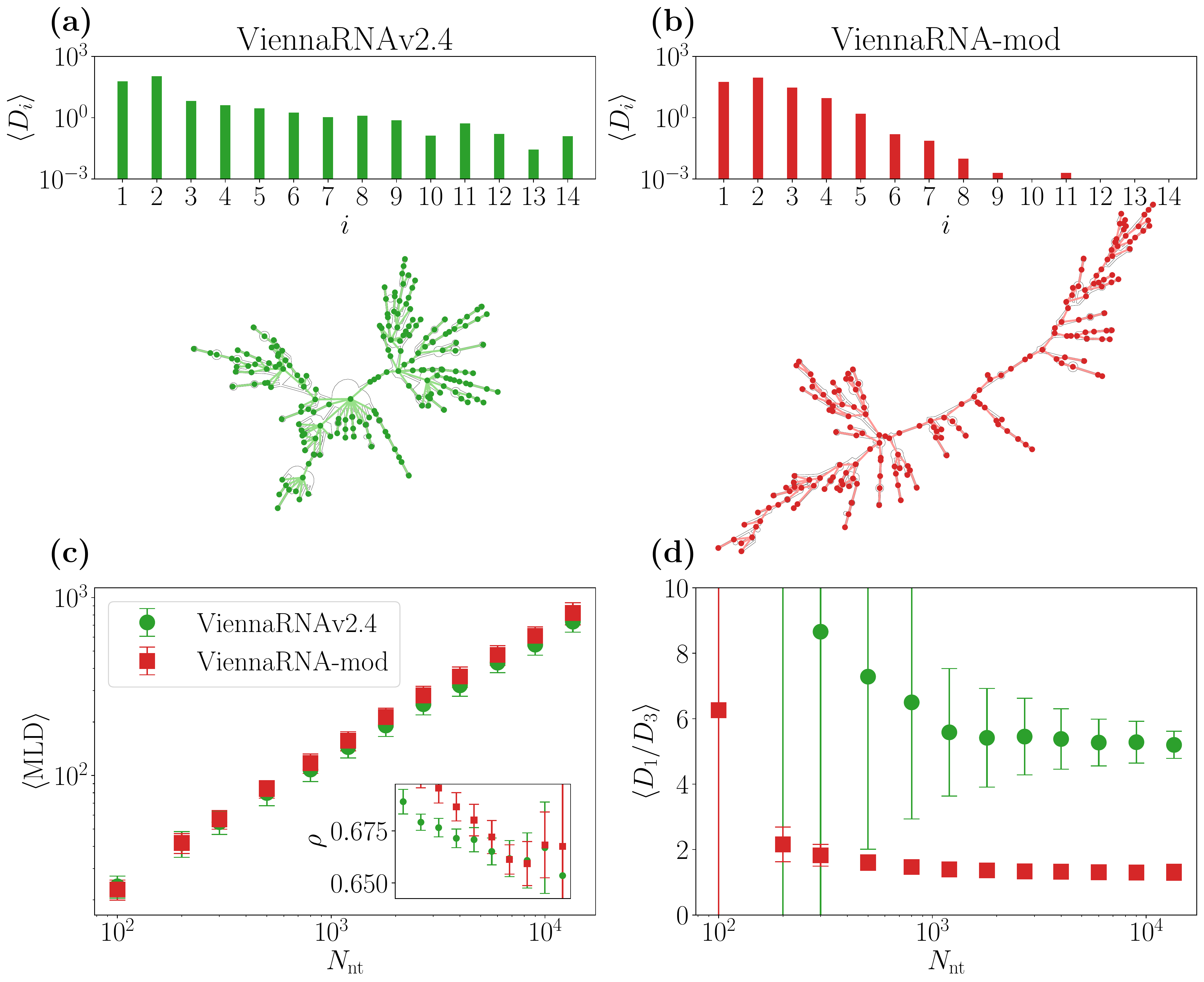}
\caption{Number of nodes of different degrees $\langle D_i\rangle$ in the BlShV genome obtained using the energy model in ViennaRNA (v2.4) with either {\bf (a)} default multiloop energy parameters or {\bf (b)} multiloop energy parameters from older versions of ViennaRNA ($<$~v2.0). For the values of these parameters, see Table~\ref{tab:multiloop}. Note the logarithmic scale in the histograms. Each panel also shows an example structure of the genome. Scaling of {\bf (c)} $\MLD$ and {\bf (d)} the ratio of the number of degree $1$ and degree $3$ nodes $D_1/D_3$ with the sequence length of uniformly random RNA as predicted by the two different sets of multiloop energy parameters.}
\label{fig:models}
\end{figure}

These differences will of course reflect in the predicted structures of long RNAs and their topological properties. Since the different energy models in Table~\ref{tab:multiloop} also differ in other aspects, it is easiest to compare multiloop energy parameters by modifying {\em only} the multiloop parameters in the current parameter set used by ViennaRNA (v2.4) with the ones from older versions ($<$~v2.0), resulting in a modified set of parameters ViennaRNA-mod. Differences in their predictions are illustrated in Fig.~\ref{fig:models}. The opposite signs of the parameter $E_\mathrm{br}$ clearly lead to very different distributions of node degrees in the genome of BlShV, with the modified set of parameters predicting far fewer multiloops ($D_{\geqslant3}$). Interestingly enough, however, the different multiloop energy parameters do not seem to lead to a different scaling behaviour of the $\MLD$ of uniformly random RNA (Fig.~\ref{fig:models}c), as the exponent $\rho$ becomes indistinguishable between the two cases in the asymptotic limit of long sequences. On the other hand, the ratio of the number of degree $1$ and degree $3$ nodes is completely different (Fig.~\ref{fig:models}d). The choice of the multiloop energy parameters can thus lead to important differences in the predicted RNA topology~\cite{Ward2017,Poznanovic2020}, which needs to be taken into account when comparing results obtained by existing studies on the branching properties of viral RNAs~\cite{Gopal2014,Borodavka2016,Yoffe2008,Tubiana2015} that use different versions of folding software and thus potentially different energy models.

\subsection{Maximum base pair span}

Parts of the folding process in very long RNA molecules (i.e., over several hundred nucleotides in length) are influenced by various factors such as co-transcriptional folding and the presence of other molecules in the cell~\cite{Amman2013,Pyle2016}. Consequently, the accuracy of RNA secondary structure prediction in general decreases with the span of a base pair---the length of the nucleotide sequence between two paired bases~\cite{Lorenz2020}. This effect can be incorporated in the folding prediction by restricting the maximum allowed base pair span~\cite{Lorenz2011,Lorenz2020}, which not only tends to yield more plausible local structure predictions but also drastically increases the computational efficiency. While restrictions on the maximum base pair span in the range of $200$--$600$ nt are often made to improve the prediction of local structural elements~\cite{Archer2013,Lan2022}, this can neglect long-range base pairing and global structure which have been shown to be important in numerous +ssRNA genomes, including that of SARS-CoV-2~\cite{Nicholson2015,Simmonds2004,Cao2021}.

\begin{figure}[t]
\centering
\includegraphics[width=0.95\linewidth]{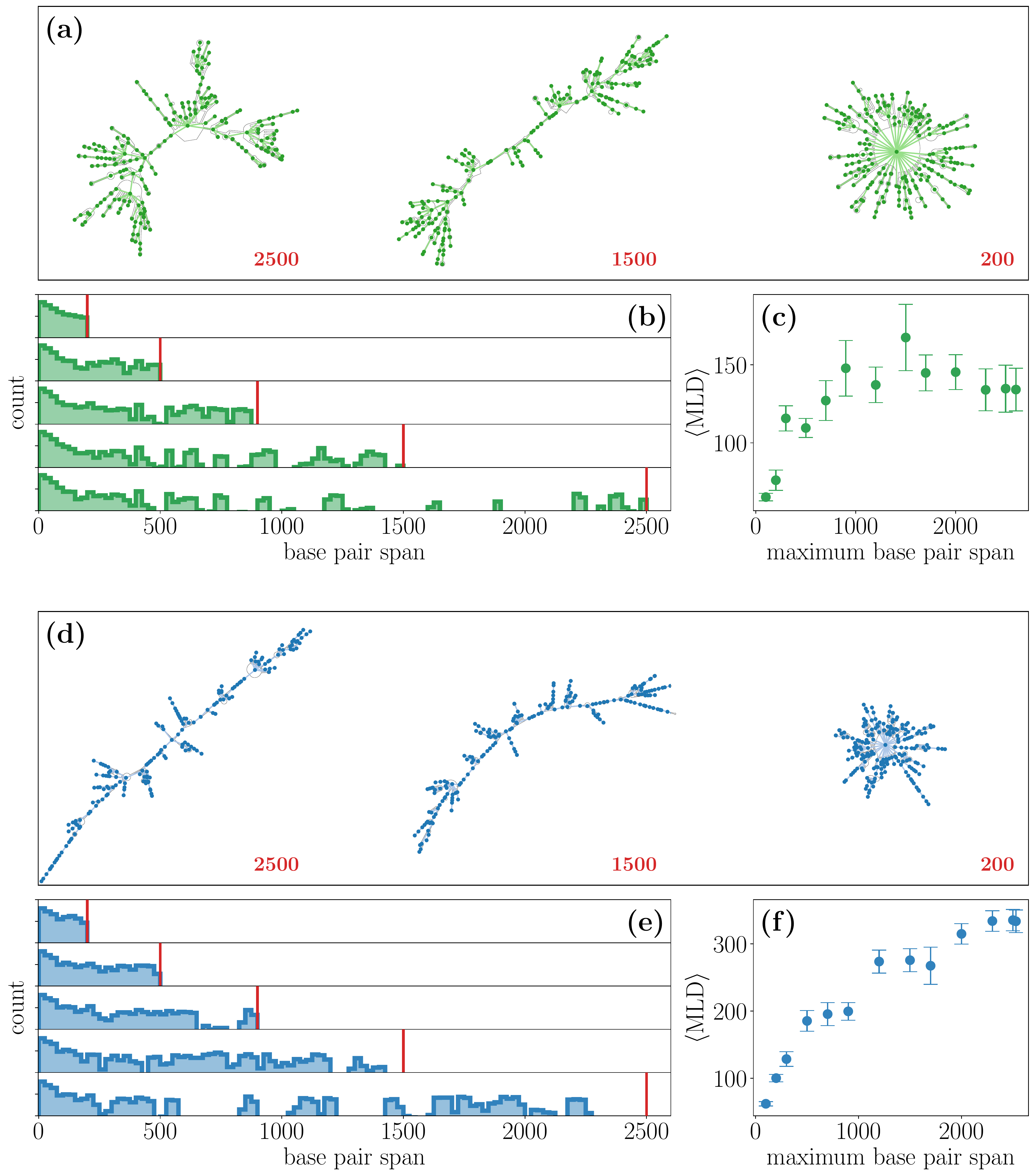}
\caption{{\bf (a)} Representative structures of BlShV genome with different maximum base pair restrictions, indicated with each structure. {\bf (b)} Distributions of base pair spans in the thermal ensemble of BlShV genome structures with different maximum base pair span restrictions (marked by vertical bars). {\bf (c)} $\MLD$ of the BlShV genome as a function of the maximum base pair span restriction. {\bf (d)--(f)} Same as panels (a)--(c) but for the BVQ genome.}
\label{fig:mbp}
\end{figure}

It is therefore natural to wonder to what extent restricting the maximum allowed base pair span influences the topological properties of the structure of viral RNAs. As panels (c) and (f) of Fig.~\ref{fig:mbp} show, restricting the maximum base pair span to $\sim1000$--$2000$ nt slightly changes the $\MLD$, regardless of whether the original genome belongs to a compact class or not, as long-range base pairs are progressively removed from the global structure (panels (b) and (e)). However, when the maximum base pair span is restricted to $\lesssim1000$ nt, $\MLD$ starts to decrease drastically. Inspection of the resulting structures (panels (a) and (d)) shows that significant maximum base pair span restrictions eventually result in RNA topology becoming more star-like because the individual hairpins are effectively being ``strung'' on a backbone of single-stranded RNA. This leads to a decrease in $\MLD$ which by definition does not take into account single-stranded regions of RNA (Sec.~\ref{sec:graph}), and could consequently affect the results showing that viral RNAs are more compact than random ones (Sec.~\ref{sec:rnaVR}). From a topological and structural perspective, imposing a drastic restriction on the maximum base pair span in RNA would thus not only require a redefinition of the MLD but perhaps even a different mapping of its secondary structure onto a graph.

\section{Field-theoretical description of viral RNA as a branched polymer}
\label{sec:ftrna}

RNA branching is also intimately connected with long-range interactions, such as electrostatic self-interaction and interactions between the RNA and capsid proteins. Since the strength of RNA self-interaction (base-pairing) is relatively weak and may easily be affected by either thermal fluctuations or electrostatic interactions~\cite{Zandi2020}, {\em annealed branched polymers} present a viable {\em coarse-grained} model system. Here, one starts with the grand canonical partition function~\cite{Lubensky1979} 
\begin{align}
\Xi(K,f_e,f_b;V)= \sum_{N, N_e, N_b} K^{N}f_e^{N_e}f_b^{N_b}\Omega(N,N_e,N_b;V)
\end{align}
where $K$ are bonds (edges), and $f_e$ and $f_b$ are end- and branch point fugacities of the annealed polymer (with hairpins counted as the end points, cf.\ Sec.~\ref{sec:graph}). Branch points (nodes) of high degree ($d_i>3)$ can be considered as combinations of branch points of degree $3$ and thus need not be treated separately in this description. The function $\Omega(N,N_e,N_b;V)$ is the number of ways to arrange $N$ bonds, $N_e$ end points, and $N_b$ branch points on a lattice of volume $V$.

The grand canonical partition function can be obtained in the $n \rightarrow 0$ limit of the partition function of an ${\cal O}(n)$ model of a magnet~\cite{deGennes1972} as a functional integral over a continuous field $\Psi$~\cite{Zandi2015}
\begin{align}
\Xi(K,f_e,f_b;V) \simeq \int {\cal D}[\Psi] e^{- \beta F_0[\psi]},
\end{align}
with the square of $\Psi$ proportional to the monomer density. The saddle-point (mean-field) free energy $F_0$---in absence of electrostatic interactions---is then 
\begin{align} \label{W_branched}
  \beta F_0[\Psi] = \!\!\int_V\!\! {\mathrm{d}^3}{{{r}}} \left[ \frac{a^2}{6} |{\nabla\Psi}|^2 + \frac{1}{2}\upsilon \Psi^4 -\frac{1}{\sqrt{a^3}}\left(f_e\Psi+\frac{a^3}{6} f_b \Psi^3\right)\right],
\end{align}
where $a$ is the statistical step length (Kuhn length; for RNA, $\sim 1$~nm)  and $\upsilon$ is the (repulsive) short-range excluded volume interaction term. Branch point and end point terms proportional to $\Psi$ and $\Psi^3$ are negative (attractive), therefore increasing the local monomer density. The annealed numbers of end- ($N_e$) and branch points ($N_b$) of the RNA (corresponding to $D_1$ and $D_{\geqslant3}$ in the graph representation, respectively) are related to the fugacities $f_e$ and $f_b$ by~\cite{Zandi2015}
\begin{align}\label{NeNb}
N_e =-  f_e \frac{\partial\beta{F}_0}{\partial{f_e}} \qquad {\rm and} \qquad N_b =-  f_b \frac{\partial\beta{F}_0}{\partial{f_b}}.
\end{align}
If the total number of monomers is fixed, the number of end points for a single RNA molecule with no closed loops depends on the number of branch points as $N_e = N_b+2$, meaning that $f_e$ is not a free parameter. The polymer is linear if $f_b=0$, and the number of branch points increases with $f_b$. 

On this level of the description of RNA branching, {\em long-range interactions} can be straightforwardly implemented by adding additional terms to the free energy functional (Eq.~\eqref{W_branched}), depending on local interaction fields and couplings with the $\Psi$ field~\cite{Erdemci2014,Erdemci2016,Siyu2017}. The free energy for, e.g., electrostatic interactions described by electrostatic potential field $\Phi$ would read
\begin{align} \label{W_electro}
  \beta F = \beta F_0[\Psi] + \beta F[\Phi] + \beta \tau \!\!\int_V\!\! {\mathrm{d}^3}{{{r}}} ~\Phi \Psi^2\, ,   
\end{align}
with $\tau$ being the charge per monomer and $\beta F[\Phi]$ the electrostatic interaction term~\cite{Siber2008}, which can also describe the interactions between the genome and the capsid proteins. On the other hand, non-electrostatic interactions implied by {\em packaging signals} require a modified approach~\cite{Man2021}.

Modifying the topology of the genome by varying the fugacities $f_e$ and $f_b$ allows to use this methodology to simulate systems of viral RNAs with a fixed number of end- and branch points. This coarse-grained description of branching is complementary to the methodology based on explicit planar tree structure~\cite{Farrell2022} and can, for instance,  differentiate between the encapsulation behavior of RNA1 of brome mosaic virus (BMV) that has $65$ branch points and of RNA1 of cowpea chlorotic mottle virus (CCMV) with $60.5$ branch points---correctly confirming that BMV RNA1 is preferentially packaged over CCMV RNA1 by the CCMV capsid protein~\cite{Erdemci2014}. Furthermore, the straightforward implementation of electrostatic interactions on top of RNA topology is probably the most important forte of this methodology and allows one to assess the role of electrostatics in spontaneous co-assembly of the negatively charged genome and positively charged capsid proteins. Using this approach, it has been demonstrated that branching in fact allows viruses to maximize the amount of encapsulated genome and makes assembly more efficient~\cite{Erdemci2014}, implies negative osmotic pressures across the capsid wall~\cite{Erdemci2016}, and can explain the effect of number and location of charges in the capsid protein tails~\cite{Dong2020}.

While the field-theoretical annealed-branching description is without doubt heavily coarse-grained, it not only provides an approximate implementation of topology, but also readily incorporates {\em short- and long-range interactions} on a level amenable to analytical calculations. Different approaches of standard polymer theory can then be transplanted into the statistical mechanics of RNA providing further insight into the coupling between topology and virion self-assembly.  

\section{Conclusions}

Mapping RNA secondary structure onto a graph enables its description as a branched polymer and a subsequent study of its topological properties (e.g., MLD and node degree distribution; Sec.~\ref{sec:RNAbranch}). This, in turn, can be connected to the physical properties of the RNA, such as its size as given by its radius of gyration. This approach also provides insight into how RNA structure compares to other types of branched polymers in terms of, for instance, their scaling exponents. Branching properties of RNA also allow comparison of +ssRNA genomes of different viral families, both among themselves and with random RNA of similar length and composition (Sec.~\ref{sec:rnaVR}). Such an analysis reveals the unusual compactness of genomes from certain viral families and may eventually provide an answer to the question of where in the sequence of viral RNA its physical compactness is encoded.

Describing RNA as a graph and in this way treating it as a branched polymer is, of course, an approximation. This description is made on the topological level of the RNA secondary structure, itself deriving from an energy-based base pair prediction, and thus depends on the model parameters used in it (Sec.~\ref{sec:models}). Secondary structure prediction furthermore remains agnostic to steric interactions between different parts of the RNA, long-range interactions such as electrostatics, and other tertiary interactions. Some of these effects can, however, be included on a coarse-grained level by treating the branching RNA structure with a field-theoretical description, enabling, for instance, a coupling between topological parameters (such as node degree distribution) and electrostatic interactions (Sec.~\ref{sec:ftrna}).

Predictions obtained by treating viral RNA as a branched polymer can, to an extent, be verified experimentally, for instance by measuring its radius of gyration by gel electrophoresis~\cite{Gopal2014} or by determining the distributions of node degrees and segment lengths from 2D projections of viral RNA molecules imaged by cryo-EM~\cite{Gopal2014,Garmann2015}. Properly designed experiments on long RNA molecules of different lengths could thus, in principle, be compared with the predictions given by different multiloop energy models and in this way help determine the most appropriate model. Lastly, treating RNA as a graph and being able to understand how its sequence leads to its topological and structural properties can be beneficial not only in the ability to interfere with the function of viral genomes but also in the design of RNA molecules with specific topology for use in nanomedicine and synthetic biology~\cite{Schlick2018,Jain2020,Geary2021}.

\begin{acknowledgement}
A.B.\ acknowledges support by Slovenian Research Agency (ARRS) under Contract No.\ P1-0055. L.T.\ acknowledges support by MIUR through the Rita Levi Montalcini grant and financial support from ICSC---Centro Nazionale di Ricerca in High Performance Computing, Big Data and Quantum Computing, funded by European Union---NextGenerationEU. R.P.\ acknowledges support from the Key Project No.\ 12034019 of the Natural Science Foundation of China. R.P.\ also thanks J.D.\ Farrell for his comments on an earlier version of the manuscript. The authors acknowledge networking support by the the COST Action No.\ CA17139 (EUTOPIA).
\end{acknowledgement}

\bibliographystyle{spphys}
\bibliography{references}

\begin{thebibliography}{100}
\providecommand{\url}[1]{{#1}}
\providecommand{\urlprefix}{URL }
\expandafter\ifx\csname urlstyle\endcsname\relax
  \providecommand{\doi}[1]{DOI \discretionary{}{}{}#1}\else
  \providecommand{\doi}{DOI \discretionary{}{}{}\begingroup
  \urlstyle{rm}\Url}\fi

\bibitem{Eddy2001}
S.R. Eddy, Nat. Rev. Genet. \textbf{2}, 919 (2001)

\bibitem{Mattick2006}
J.S. Mattick, I.V. Makunin, Human Mol. Genet. \textbf{15}, R17 (2006)

\bibitem{Gorodkin2014}
J.~Gorodkin, W.L. Ruzzo, \emph{{RNA} sequence, structure, and function:
  Computational and bioinformatic methods} (Springer, 2014)

\bibitem{Wang2021}
X.W. Wang, C.X. Liu, L.L. Chen, Q.C. Zhang, Nat. Chem. Biol. \textbf{17}, 755
  (2021)

\bibitem{Mortimer2014}
S.A. Mortimer, M.A. Kidwell, J.A. Doudna, Nat. Rev. Genet. \textbf{15}, 469
  (2014)

\bibitem{Brion1997}
P.~Brion, E.~Westhof, Annu. Rev. Biophys. Biomol. Struct. \textbf{26}, 113
  (1997)

\bibitem{Mustoe2014}
A.M. Mustoe, C.L. Brooks, H.M. Al-Hashimi, Annu. Rev. Biochem. \textbf{83}, 441
  (2014)

\bibitem{Leontis2012}
N.~Leontis, E.~Westhof (eds.), \emph{{RNA 3D} structure analysis and
  prediction} (Springer, 2012)

\bibitem{Miao2017}
Z.~Miao, E.~Westhof, Annu. Rev. Biophys. \textbf{46}, 483 (2017)

\bibitem{Li2021}
J.~Li, S.J. Chen, Front. Mol. Biosci. \textbf{8} (2021)

\bibitem{Low2010}
J.T. Low, K.M. Weeks, Methods \textbf{52}, 150 (2010)

\bibitem{Lorenz2016}
R.~Lorenz, I.L. Hofacker, P.F. Stadler, Algorithms Mol. Biol. \textbf{11}, 1
  (2016)

\bibitem{Mitchell2019}
D.~Mitchell~III, S.M. Assmann, P.C. Bevilacqua, Curr. Op. Struct. Biol.
  \textbf{59}, 151 (2019)

\bibitem{Holmes2009}
E.C. Holmes, \emph{The evolution and emergence of {RNA} viruses} (Oxford
  University Press, 2009)

\bibitem{Liu2009}
Y.~Liu, E.~Wimmer, A.V. Paul, Biochimi. Biophys. Acta \textbf{1789}, 495 (2009)

\bibitem{Newburn2015}
L.R. Newburn, K.A. White, Virology \textbf{479}, 434 (2015)

\bibitem{Nicholson2015}
B.L. Nicholson, K.A. White, Curr. Op. Virol. \textbf{12}, 66 (2015)

\bibitem{Boerneke2019}
M.A. Boerneke, J.E. Ehrhardt, K.M. Weeks, Annu. Rev. Virol. \textbf{6}, 93
  (2019)

\bibitem{Schneemann2006}
A.~Schneemann, Annu. Rev. Microbiol. \textbf{60}, 51 (2006)

\bibitem{Rao2006}
A.~Rao, Annu. Rev. Phytopathol. \textbf{44}, 61 (2006)

\bibitem{Garmann2016}
R.F. Garmann, M.~Comas-Garcia, C.M. Knobler, W.M. Gelbart, Acc. Chem. Res.
  \textbf{49}, 48 (2016)

\bibitem{ComasGarcia2019}
M.~Comas-Garcia, Viruses \textbf{11}, 253 (2019)

\bibitem{Twarock2018}
R.~Twarock, R.J. Bingham, E.C. Dykeman, P.G. Stockley, Curr. Op. Virol.
  \textbf{31}, 74 (2018)

\bibitem{Stockley2013}
P.G. Stockley, R.~Twarock, S.E. Bakker, A.M. Barker, A.~Borodavka, E.~Dykeman,
  R.J. Ford, A.R. Pearson, S.E. Phillips, N.A. Ranson, et~al., J. Biol. Phys.
  \textbf{39}, 277 (2013)

\bibitem{Zandi2020}
R.~Zandi, B.~Dragnea, A.~Travesset, R.~Podgornik, Phys. Rep. \textbf{847}, 1
  (2020)

\bibitem{Perlmutter2015}
J.D. Perlmutter, M.F. Hagan, Annu. Rev. Phys. Chem. \textbf{66}, 217 (2015)

\bibitem{Hu2008}
Y.~Hu, R.~Zandi, A.~Anavitarte, C.M. Knobler, W.M. Gelbart, Biophys. J.
  \textbf{94}, 1428 (2008)

\bibitem{ComasGarcia2012}
M.~Comas-Garcia, R.D. Cadena-Nava, A.~Rao, C.M. Knobler, W.M. Gelbart, J.
  Virol. \textbf{86}, 12271 (2012)

\bibitem{Beren2017}
C.~Beren, L.L. Dreesens, K.N. Liu, C.M. Knobler, W.M. Gelbart, Biophys. J.
  \textbf{113}, 339 (2017)

\bibitem{Marichal2021}
L.~Marichal, L.~Gargowitsch, R.L. Rubim, C.~Sizun, K.~Kra, S.~Bressanelli,
  Y.~Dong, S.~Panahandeh, R.~Zandi, G.~Tresset, Biophys. J. \textbf{120}, 3925
  (2021)

\bibitem{Perlmutter2013}
J.D. Perlmutter, C.~Qiao, M.F. Hagan, eLife \textbf{2} (2013)

\bibitem{Garmann2022}
R.F. Garmann, A.M. Goldfain, C.R. Tanimoto, C.E. Beren, F.F. Vasquez, D.A.
  Villarreal, C.M. Knobler, W.M. Gelbart, V.N. Manoharan, Proc. Natl. Acad.
  Sci. USA \textbf{119}, e2206292119 (2022)

\bibitem{Poblete2021}
S.~Poblete, A.~{Bo\v{z}i\v{c}}, M.~Kandu\v{c}, R.~Podgornik, H.A. {Vargas
  Guzm\'{a}n}, ACS Omega \textbf{6}, 32823 (2021)

\bibitem{Singaram2015}
S.W. Singaram, R.F. Garmann, C.M. Knobler, W.M. Gelbart, A.~Ben-Shaul, J. Phys.
  Chem. B \textbf{119}, 13991 (2015)

\bibitem{Erdemci2014}
G.~Erdemci-Tandogan, J.~Wagner, P.~Van Der~Schoot, R.~Podgornik, R.~Zandi,
  Phys. Rev. E \textbf{89}, 032707 (2014)

\bibitem{Erdemci2016}
G.~Erdemci-Tandogan, J.~Wagner, P.~van~der Schoot, R.~Podgornik, R.~Zandi,
  Phys. Rev. E \textbf{94}, 022408 (2016)

\bibitem{Gopal2014}
A.~Gopal, D.E. Egecioglu, A.M. Yoffe, A.~Ben-Shaul, A.L. Rao, C.M. Knobler,
  W.M. Gelbart, PLoS One \textbf{9}, e105875 (2014)

\bibitem{Borodavka2016}
A.~Borodavka, S.W. Singaram, P.G. Stockley, W.M. Gelbart, A.~Ben-Shaul,
  R.~Tuma, Biophys. J. \textbf{111}, 2077 (2016)

\bibitem{Yoffe2008}
A.M. Yoffe, P.~Prinsen, A.~Gopal, C.M. Knobler, W.M. Gelbart, A.~Ben-Shaul,
  Proc. Natl. Acad. Sci. USA \textbf{105}, 16153 (2008)

\bibitem{Tubiana2015}
L.~Tubiana, A.~Bo{\v{z}}i{\v{c}}, C.~Micheletti, R.~Podgornik, Biophys. J.
  \textbf{108}, 194 (2015)

\bibitem{Bozic2018}
A.~{Bo\v{z}i\v{c}}, C.~Micheletti, R.~Podgornik, L.~Tubiana, J. Phys. Condens.
  Matter \textbf{30}, 084006 (2018)

\bibitem{Farrell2022}
J.~Farrell, J.~Dobnikar, R.~Podgornik, Phys. Rev. Res. \textbf{5}, L012040
  (2023)

\bibitem{Fallmann2017}
J.~Fallmann, S.~Will, J.~Engelhardt, B.~Gr{\"u}ning, R.~Backofen, P.F. Stadler,
  J. Biotechnol. \textbf{261}, 97 (2017)

\bibitem{Lorenz2011}
R.~Lorenz, S.H. Bernhart, C.~H{\"o}ner~zu Siederdissen, H.~Tafer, C.~Flamm,
  P.F. Stadler, I.L. Hofacker, Algorithms Mol. Biol. \textbf{6}, 1 (2011)

\bibitem{Reuter2010}
J.S. Reuter, D.H. Mathews, BMC Bioinform. \textbf{11}, 1 (2010)

\bibitem{Do2006}
C.B. Do, D.A. Woods, S.~Batzoglou, Bioinformatics \textbf{22}, e90 (2006)

\bibitem{Wayment2020}
H.K. Wayment-Steele, W.~Kladwang, A.I. Strom, J.~Lee, A.~Treuille, A.~Becka,
  {Eterna Participants}, R.~Das, Nat. Methods \textbf{19}, 1234 (2022)

\bibitem{Koodli2021}
R.V. Koodli, B.~Rudolfs, H.K. Wayment-Steele, {Eterna Structure Designers},
  R.~Das, bioRxiv  (2021).
\newblock
  \urlprefix\url{https://www.biorxiv.org/content/10.1101/2021.08.26.457839v1}

\bibitem{Liu2021}
M.~Liu, E.~Poppleton, G.~Pedrielli, P.~{\v{S}}ulc, D.P. Bertsekas, INFORMS J.
  Comput.  (2022)

\bibitem{Spasic2018}
A.~Spasic, S.M. Assmann, P.C. Bevilacqua, D.H. Mathews, Nucleic Acids Res.
  \textbf{46}, 314 (2018)

\bibitem{Mathews2006}
D.H. Mathews, D.H. Turner, Curr. Op. Struct. Biol. \textbf{16}, 270 (2006)

\bibitem{Schlick2018}
T.~Schlick, Methods \textbf{143}, 16 (2018)

\bibitem{GraphTheory}
J.L. Gross, J.~Yellen, M.~Anderson, \emph{Graph theory and its applications}
  (Chapman and Hall/CRC, 2018)

\bibitem{Todeschini2008}
R.~Todeschini, V.~Consonni, \emph{Handbook of molecular descriptors} (John
  Wiley \& Sons, 2008)

\bibitem{Rouvray2002}
D.H. Rouvray, R.B. King, \emph{Topology in chemistry: Discrete mathematics of
  molecules} (Elsevier, 2002)

\bibitem{Sazer2018}
S.~Sazer, H.~Schiessel, Traffic \textbf{19}, 87 (2018)

\bibitem{Perry2019}
S.L. Perry, Curr. Op. Colloid Interface Sci. \textbf{39}, 86 (2019)

\bibitem{Zandi2015}
J.~Wagner, G.~Erdemci-Tandogan, R.~Zandi, J. Phys.: Condens. Matter
  \textbf{27}, 495101 (2015)

\bibitem{Gutin1993}
A.M. Gutin, A.Y. Grosberg, E.I. Shakhnovich, Macromolecules \textbf{26}(6),
  1293 (1993)

\bibitem{Everaers2017}
R.~Everaers, A.Y. Grosberg, M.~Rubinstein, A.~Rosa, Soft Matter \textbf{13}(6),
  1223 (2017)

\bibitem{Wang2017}
Z.G. Wang, Macromolecules \textbf{50}(23), 9073 (2017)

\bibitem{Giacometti2013}
S.M. Bhattacharjee, A.~Giacometti, A.~Maritan, J. Phys. Cond. Matter
  \textbf{25}(50), 503101 (2013)

\bibitem{RubinsteinColbyBook}
M.~Rubinstein, R.H. Colby, \emph{Polymer Physics} (Oxford University Press, New
  York, 2003)

\bibitem{LiMadrasSokal1995}
B.~Li, N.~Madras, A.D. Sokal, J. Stat. Phys. \textbf{80}(3), 661 (1995)

\bibitem{ParisiSourlas1981}
G.~Parisi, N.~Sourlas, Phys. Rev. Lett. \textbf{46}, 871 (1981)

\bibitem{vanRensburg1992}
E.J. Van~Rensburg, N.~Madras, J. Phys. A: Math. Theor. \textbf{25}, 303 (1992)

\bibitem{RosaEveraersJPA2016}
A.~Rosa, R.~Everaers, J. Phys. A. Math. Theor. \textbf{49}, 345001 (2016)

\bibitem{RosaEveraersJCP2016}
A.~Rosa, R.~Everaers, J. Chem. Phys. \textbf{145}, 164906 (2016)

\bibitem{FloryChemBook}
P.J. Flory, \emph{Principles of Polymer Chemistry} (Cornell University Press,
  Ithaca (NY), 1953)

\bibitem{Simon2021}
D.~Sim{\'o}n, J.~Cristina, H.~Musto, Front. Microbiol. \textbf{12} (2021)

\bibitem{Schultes1997}
E.~Schultes, P.T. Hraber, T.H. LaBean, RNA \textbf{3}, 792 (1997)

\bibitem{Higgs1993}
P.G. Higgs, J. Phys., I \textbf{3}, 43 (1993)

\bibitem{Clote2005}
P.~Clote, F.~Ferr{\'e}, E.~Kranakis, D.~Krizanc, RNA \textbf{11}, 578 (2005)

\bibitem{Lefkowitz2018}
E.J. Lefkowitz, D.M. Dempsey, R.C. Hendrickson, R.J. Orton, S.G. Siddell, D.B.
  Smith, Nucleic Acids Res. \textbf{46}, D708 (2018)

\bibitem{Gaunt2022}
E.R. Gaunt, P.~Digard, Wiley Interdiscip. Rev. RNA \textbf{13}, e1679 (2022)

\bibitem{Giallonardo2017}
F.~Di~Giallonardo, T.E. Schlub, M.~Shi, E.C. Holmes, J. Virol. \textbf{91},
  e02381 (2017)

\bibitem{Belalov2013}
I.S. Belalov, A.N. Lukashev, PLOS One \textbf{8}, e56642 (2013)

\bibitem{Jiang2008}
M.~Jiang, J.~Anderson, J.~Gillespie, M.~Mayne, BMC Bioinform. \textbf{9}, 1
  (2008)

\bibitem{Zhao2018}
Y.~Zhao, J.~Wang, C.~Zeng, Y.~Xiao, Biophys. Rep. \textbf{4}, 123 (2018)

\bibitem{Turner2010}
D.H. Turner, D.H. Mathews, Nucleic Acids Res. \textbf{38}, D280 (2010)

\bibitem{Andronescu2010}
M.~Andronescu, A.~Condon, H.H. Hoos, D.H. Mathews, K.P. Murphy, RNA
  \textbf{16}, 2304 (2010)

\bibitem{Langdon2018}
W.B. Langdon, J.~Petke, R.~Lorenz, in \emph{European Conference on Genetic
  Programming} (2018), pp. 220--236

\bibitem{Poznanovic2021}
S.~Poznanovi{\'c}, C.~Wood, M.~Cloer, C.~Heitsch, Genes \textbf{12}, 469 (2021)

\bibitem{Wiedemann2022}
J.~Wiedemann, J.~Kaczor, M.~Milostan, T.~Zok, J.~Blazewicz, M.~Szachniuk,
  M.~Antczak, Bioinformatics  (2022)

\bibitem{Zuker1981}
M.~Zuker, P.~Stiegler, Nucleic Acids Res. \textbf{9}, 133 (1981)

\bibitem{Ward2017}
M.~Ward, A.~Datta, M.~Wise, D.H. Mathews, Nucleic Acids Res. \textbf{45}, 8541
  (2017)

\bibitem{Ward2019}
M.~Ward, H.~Sun, A.~Datta, M.~Wise, D.H. Mathews, Bioinformatics \textbf{35},
  4298 (2019)

\bibitem{Poznanovic2020}
S.~Poznanovi{\'c}, F.~Barrera-Cruz, A.~Kirkpatrick, M.~Ielusic, C.~Heitsch, J.
  Struct. Biol. \textbf{210}, 107475 (2020)

\bibitem{Amman2013}
F.~Amman, S.H. Bernhart, G.~Doose, I.L. Hofacker, J.~Qin, P.F. Stadler,
  S.~Will, in \emph{Brazilian Symposium on Bioinformatics} (Springer, 2013),
  pp. 1--11

\bibitem{Pyle2016}
A.M. Pyle, T.~Schlick, J. Mol. Biol. \textbf{428}, 733 (2016)

\bibitem{Lorenz2020}
R.~Lorenz, P.F. Stadler, Genes \textbf{12}, 14 (2020)

\bibitem{Archer2013}
E.J. Archer, M.A. Simpson, N.J. Watts, R.~O’Kane, B.~Wang, D.A. Erie,
  A.~McPherson, K.M. Weeks, Biochemistry \textbf{52}, 3182 (2013)

\bibitem{Lan2022}
T.C. Lan, M.F. Allan, L.E. Malsick, J.Z. Woo, C.~Zhu, F.~Zhang, S.~Khandwala,
  S.S. Nyeo, Y.~Sun, J.U. Guo, et~al., Nat. Comm. \textbf{13}, 1 (2022)

\bibitem{Simmonds2004}
P.~Simmonds, A.~Tuplin, D.J. Evans, RNA \textbf{10}, 1337 (2004)

\bibitem{Cao2021}
C.~Cao, Z.~Cai, X.~Xiao, J.~Rao, J.~Chen, N.~Hu, M.~Yang, X.~Xing, Y.~Wang,
  M.~Li, et~al., Nat. Comm. \textbf{12}, 1 (2021)

\bibitem{Lubensky1979}
T.~Lubensky, J.~Isaacson, Phys. Rev. A \textbf{20}, 2130 (1979)

\bibitem{deGennes1972}
T.~Lubensky, J.~Isaacson, Phys. Lett. A \textbf{38}, 339 (1972)

\bibitem{Siyu2017}
S.~Li, G.~Erdemci-Tandogan, J.~Wagner, P.~van~der Schoot, R.~Zandi, Phys. Rev.
  E \textbf{96}, 22401 (2017)

\bibitem{Siber2008}
A.~\v{S}iber, R.~Podgornik, Phys. Rev. E \textbf{78}, 051915 (2008)

\bibitem{Man2021}
C.~Huang, R.~Podgornik, X.~Man, Macromolecules \textbf{54}, 9602 (2021)

\bibitem{Dong2020}
Y.~Dong, S.~Li, R.~Zandi., Phys. Rev. E \textbf{102}, 062423 (2020)

\bibitem{Garmann2015}
R.F. Garmann, A.~Gopal, S.S. Athavale, C.M. Knobler, W.M. Gelbart, S.C. Harvey,
  RNA \textbf{21}, 877 (2015)

\bibitem{Jain2020}
S.~Jain, Y.~Tao, T.~Schlick, J. Struct. Biol. \textbf{209}, 107438 (2020)

\bibitem{Geary2021}
C.~Geary, G.~Grossi, E.K. McRae, P.W. Rothemund, E.S. Andersen, Nat. Chem.
  \textbf{13}, 549 (2021)

\end{thebibliography}

\end{document}